\documentclass[11pt]{article}
\usepackage{epsfig} 
\usepackage{graphics}
\newcommand{\sect}[1]{ \section{#1} \setcounter{equation}{0} }

\newcommand{\Nf}{N_{\!f}}
\newcommand{\NA}{N_{\!A}}
\newcommand{\NF}{N_{\!F}}
\newcommand{\bare}{\mbox{\footnotesize{o}}}
\newcommand{\Dslash}{D \! \! \! \! /}

\newcommand{\pslash}{p \! \! \! /}
\newcommand{\MSbar}{\overline{\mbox{MS}}}
\newcommand{\MSbars}{\overline{\mbox{\footnotesize{MS}}}}
\newcommand{\RI}{\mbox{RI${}^\prime$}}
\newcommand{\RIs}{\mbox{\footnotesize{RI${}^\prime$}}}

\begin{document}

\title{Five loop anomalous dimension of non-singlet quark currents in the
RI${}^\prime$ scheme}

\author{J.A. Gracey, \\ Theoretical Physics Division, \\ 
Department of Mathematical Sciences, \\ University of Liverpool, \\ P.O. Box 
147, \\ Liverpool, \\ L69 3BX, \\ United Kingdom.} 
\date{}

\maketitle 

\vspace{5cm} 
\noindent 
{\bf Abstract.} We construct the five loop anomalous dimensions of the basic
fields in Quantum Chromodynamics in a linear covariant gauge in the modified 
Regularization Invariant ($\RI$) scheme. Using these core results we also 
compute the four loop Green's function where the quark mass operator and vector
current are separately inserted in a quark $2$-point function. These are 
necessary for measurements of the same quantities on the lattice. The Green's 
functions are provided in both the modified Minimal Subtraction ($\MSbar$) and 
$\RI$ schemes. All the results are available for a general Lie group.

\vspace{-17.2cm}
\hspace{13.2cm}
{\bf LTH 1317}

\newpage 

\sect{Introduction.}

Using lattice gauge theory to study the non-perturbative regime of Quantum
Chromodynamics (QCD) has advanced our understanding of quark masses as well as
providing accurate theoretical predictions of masses such as that of the proton
and neutron together with the spectrum of mesons and other baryons. Central to 
lattice gauge theory analyses is the use of supercomputers to numerically 
evaluate Green's functions, for instance, in the low energy region where the
mechanisms behind colour and quark confinement operate. Measurements of such
Green's functions are carried out on a regular discretized spacetime with the
separation distance acting as a regulator that when reduced in magnitude 
requires a careful extrapolation to continuum spacetime. This has to be 
implemented in such a way that errors are under control and not significant. 
Such a task is not straightforward and while this problem is generally very 
well understood now, having independent information on any Green's function of 
interest would provide an important quality control tool. Indeed it was
recognized in \cite{1,2}, for instance, and emphasised in \cite{3} that the
high loop order perturbative evaluation of Green's functions could provide 
useful information, albeit at high energy. In other words a series of low 
energy lattice measurements could be extrapolated to larger energies which 
should therefore overlap with a similar extension of the perturbative result to
lower energies. Such a programme has proved successful over the years in 
contributing to the understanding and improvement of lattice errors. It has 
also benefitted from the intense use of high performance computing (HPC) 
facilities devoted to lattice computations as well as the development of 
techniques to evaluate Feynman graphs to higher loop order using new algorithms
written in efficient symbolic manipulation languages. These have equally 
gained from similar improvements in computing power.

While this summarizes the overall vision of the earlier ideas of \cite{1,2,3} 
there are ongoing technical and field theory issues to be overcome. On the 
lattice side evaluating Green's functions can be financially expensive 
especially when they involve operators with a spacetime derivative. This can be
circumvented to a degree by using a specific renormalization scheme developed 
in \cite{1,2} and named the modified regularization invariant ($\RI$) scheme. 
It is a minor variation of the related RI scheme but $\RI$ is more popularly 
used now since it does not require taking a spacetime derivative on the 
lattice. By contrast the scheme that is used in virtually all perturbative 
continuum calculations is the modified minimal subtraction ($\MSbar$) scheme, 
\cite{4,5}. It is usually coupled with a regularization that is dimensional 
where the spacetime dimension is analytically continued to 
$d$~$=$~$4$~$-$~$2\epsilon$ with the small parameter $\epsilon$ acting as the 
regularizing parameter. The main reason why the $\MSbar$ scheme is used 
together with dimensional regularization is that in this configuration one can 
evaluate massless Feynman integrals and thence renormalize Green's functions to
a very high loop order. A case in point is the five loop renormalization of 
QCD, \cite{6,7,8,9,10}. While the lattice measurements of Green's functions are
carried out in the $\RI$ scheme the parallel continuum calculations are always 
in $\MSbar$. Therefore before matching comparisons and subsequent error 
analyses can be determined the dependence of the Green's function over all
energy scales has to be found in the same scheme. For practical reasons this is
much easier to carry out in the continuum since the $\RI$ scheme can be defined
in that case, \cite{1,2,11,12,13}, allowing for the renormalization of QCD. 
However, with the move to improve precision on the lattice low energy side 
there is now a need to progress loop calculations beyond the present orders 
that are available for Green's functions that are of immediate importance. 

This is the purpose of the article. We will renormalize the QCD Lagrangian at 
four loops in the $\RI$ scheme and by using properties of the renormalization 
group we will determine the renormalization group functions of the fields at 
{\em five} loops. The $\RI$ $\beta$-function is already trivially available 
since the scheme is defined in such a way that the coupling constant 
renormalization is carried out in an $\MSbar$ way. In other words at the 
subtraction point of a vertex function only the poles with respect to 
$\epsilon$ are absorbed into the coupling renormalization constant. Therefore 
the coefficients of each term of the five loop $\RI$ scheme $\beta$-function 
are formally equivalent to those of the $\MSbar$ one. The reason why the field 
anomalous dimensions are important in the $\RI$ scheme is that they are 
required for renormalizing Green's functions with operator insertions, 
\cite{1,2}. In particular recent lattice studies \cite{14,15,16} have, for 
instance, measured quark bilinear operators inserted in a quark $2$-point 
function at zero momentum in the $\RI$ scheme. So as such Green's functions are
of current interest we have also renormalized the quark mass operator and 
vector current to four loops and will provide the analytic and numerical values
of the respective Green's functions to the same order. For example the former 
will be useful for lattice studies that relate to quark mass determination. The
article therefore extends the three loop results of \cite{11,12,13}. One 
corollary is that we will provide the five loop $\RI$ quark mass dimension for 
a general colour group in a linear covariant gauge. Although lattice 
measurements are invariably in the Landau gauge the continuum calculations are 
carried out for a non-zero gauge parameter partly as it can be used for 
cross-checks on the evaluation of the Feynman graphs. However it is also 
important to have such data for other problems. For instance, the study of the 
conformal window in gauge theories has been an important pursuit in recent 
years given the potential for it to give insight into four dimensional 
conformal field theories, as well as the connection with beyond the Standard 
Model ideas. The conformal window is governed by the existence of the 
Banks-Zaks fixed point \cite{17,18}. Such a fixed point has associated critical
exponents that can be estimated perturbatively. As they are renormalization 
group invariants they have to be scheme independent although in a truncation of
their perturbative expansion this would only be true approximately. So having 
the $\RI$ scheme renormalization group functions to a new level of precision 
will be important for extending the results of \cite{19,20,21} which is 
currently in progress, \cite{22}. As an example the measurement of the quark 
mass anomalous dimension exponent has been looked at on the lattice in 
\cite{23,24}. Having the five loop result in another scheme could assist with 
improving the continuum estimate.

The article is organized as follows. We discuss the method for extracting the 
four loop $\RI$ renormalization constants in Section $2$ as well as how we use
properties of the renormalization group to deduce the five loop anomalous
dimensions of the fields and covariant gauge parameter in the same scheme. The
corresponding results are contained in Section $3$ together with the related
conversion functions. Sections $4$ and $5$ respectively focus on the four loop 
renormalization of the quark mass operator and vector current. They contain the
value of the Green's function under consideration in both the $\MSbar$ and 
$\RI$ schemes. We summarize our conclusions in Section $6$ while Appendix A 
records expressions for a general Lie group.

\sect{Background.}

By way of orientation we recall the bare QCD Lagrangian with a linear covariant
gauge fixing is
\begin{equation}
L ~=~ -~ \frac{1}{4} G_{\bare \, \mu\nu}^a G_{\bare}^{a \, \mu\nu} ~-~ 
\frac{1}{2\alpha_{\bare}} (\partial^\mu A^a_{\bare \, \mu} )^2 ~-~ 
\bar{c}_{\bare}^a \partial_\mu D_{\bare}^\mu c_{\bare}^a ~+~
i \bar{\psi}_{\bare}^{iI} \Dslash_{\bare} \psi_{\bare}^{iI}
\label{lag}
\end{equation}
where the subscript ${}_{\bare}$ denotes a bare field or parameter. The 
renormalization constants are introduced in the canonical way through a 
rescaling of the bare entities
\begin{equation}
A^{a \, \mu}_{\mbox{\footnotesize{o}}} ~=~ \sqrt{Z_A} \, A^{a \, \mu} ~~,~~
c^a_{\mbox{\footnotesize{o}}} ~=~ \sqrt{Z_c} \, c^a ~~,~~
\psi_{\mbox{\footnotesize{o}}} ~=~ \sqrt{Z_\psi} \psi ~~,~~
g_{\mbox{\footnotesize{o}}} ~=~ \mu^\epsilon Z_g \, g ~~,~~
\alpha_{\mbox{\footnotesize{o}}} ~=~ Z^{-1}_\alpha Z_A \, \alpha
\label{zdef}
\end{equation}
where $g$ is the coupling constant, $\mu$ is the arbitrary mass scale 
associated with the dimensionally regularized version of the action. It is 
these renormalization constants that are determined with respect to a scheme, 
and in particular the $\RI$ scheme, that we will be interested in here. We 
recall that the $\RI$ scheme was introduced in connection with the 
renormalization of operators of interest to lattice field theory but can be 
equally applied to the renormalization of the Lagrangian in the continuum, 
\cite{1,2}. This was carried out in \cite{11,12,13} to three loops previously. 
Briefly the prescription defining the $\RI$ scheme is that the renormalization 
constants of the fields are fixed by ensuring that at the subtraction point 
each $2$-point function has no $O(a)$ correction where $a$~$=$~$g^2/(16\pi^2)$.
While this is similar to the momentum subtraction scheme given in \cite{25,26} 
it differs in the prescription for defining the coupling renormalization 
constant. In that case $Z_g$ is determined using the same criterion as that of 
the $\MSbar$ scheme. Consequently the $\RI$ scheme $\beta$-function is formally
the same as that of the $\MSbar$ one in terms of the coefficients of the 
polynomial in $a$ but that variable is regarded as an $\RI$ scheme coupling 
constant rather than an $\MSbar$ one. 
 
In \cite{13} the three loop renormalization constants were deduced from 
applying the {\sc Mincer} algorithm, \cite{27,28}, to the evaluation of the 
Feynman graphs comprising the gluon, ghost and quark $2$-point functions. While
QCD had been renormalized at four loops, \cite{29,30,31,32,33}, this was 
achieved via a variety of different techniques. More recently though the
renormalization of QCD has progressed to the five loop level,
\cite{6,7,8,9,10,33,34,35}. Underlying the computations carried out by one
group was the extension of {\sc Mincer} to its four loop successor 
{\sc Forcer}, \cite{36,37}, together with infrared rearrangement, \cite{38,39}.
One consequence of the use of {\sc Forcer} was the provision of each of the 
$2$-point functions to four loops in terms of bare parameters to particularly 
high order in the $\epsilon$ expansion, \cite{40}. This data provides the core 
starting point for implementing the renormalization of (\ref{lag}) in the $\RI$
scheme. Therefore we have applied the $\RI$ prescription using the symbolic 
manipulation language {\sc Form}, \cite{41,42}, and extracted the four loop 
values of $Z_A$, $Z_\alpha$, $Z_c$ and $Z_\psi$ using the method of \cite{43}. 
These are converted into the associated renormalization group functions via the
relations
\begin{eqnarray}
\gamma^{\RIs}_\phi(a,\alpha) &=& 
\beta^{\RIs}(a) \frac{\partial \ln Z^{\RIs}_\phi}{\partial a} ~+~ 
\alpha \gamma^{\RIs}_\alpha(a,\alpha) 
\frac{\partial \ln Z^{\RIs}_\phi}{\partial \alpha} \nonumber \\
\gamma^{\RIs}_\alpha(a,\alpha) &=& 
\left[ \beta^{\RIs}(a) \frac{\partial \ln Z^{\RIs}_\alpha}{\partial a} ~-~ 
\gamma^{\RIs}_A(a,\alpha) \right]
\left[ 1 ~-~ \alpha \frac{\partial \ln Z^{\RIs}_\alpha}{\partial \alpha} 
\right]^{-1}
\end{eqnarray}
where $\phi$ represents either gluon, ghost or quark with respective labels
$A$, $c$ and $\psi$ for future reference. The relation for the gauge parameter
$\alpha$ is for a covariant gauge fixing parameter in general but in the case
of the linear gauge fixing used here $Z_\alpha$ is unity in any scheme leading 
to the relation
\begin{equation}
\gamma^{\mbox{\footnotesize{RI$^\prime$}}}_A(a,\alpha) ~=~ -~
\gamma^{\mbox{\footnotesize{RI$^\prime$}}}_\alpha(a,\alpha) 
\end{equation}
which is true to all orders. We use the convention that the variables of the
renormalization group functions are those in the scheme indicated in the
superscript unless they are labelled explicitly. Also the argument of the $\RI$
$\beta$-function does not involve $\alpha$ since the vertex renormalization is 
carried out according to the $\MSbar$ prescription.

We delay presenting the explicit four loop expressions for the simple reason
that it is possible to determine the five loop renormalization group functions 
in the $\RI$ scheme. This is achieved from knowledge of several quantities
which are the five loop $\MSbar$ renormalization group functions from 
\cite{9,10,33,34,35} and the full four loop renormalization constants in the 
$\RI$ scheme itself. By the latter we mean that the finite part of the
renormalization constants is absolutely crucial to extending our four loop 
$\RI$ scheme expressions to the next loop order. These allow us to construct
conversion functions which provide the second key ingredient of the relevant
formalism. From the properties of the renormalization group the anomalous
dimensions in separate schemes are related by
\begin{eqnarray}
\gamma^{\RIs}_\phi \left( a_{\RIs}, \alpha_{\RIs} \right) &=& \left[
\gamma^{\MSbars}_\phi \left( a_{\MSbars}, \alpha_{\MSbars} \right) ~+~
\beta^{\MSbars} \left( a_{\MSbars} \right) 
\frac{\partial ~}{\partial a_{\MSbars}}
\ln C_\phi \left( a_{\MSbars}, \alpha_{\MSbars} \right) \right. \nonumber \\
&& \left. ~+~ \alpha_{\MSbars} \, \gamma^{\MSbars}_\alpha \left( a_{\MSbars},
\alpha_{\MSbars} \right) \frac{\partial ~}{\partial \alpha_{\MSbars}}
\ln C_\phi \left( a_{\MSbars}, \alpha_{\MSbars} \right)
\right]_{\MSbars \to \RIs} ~.
\label{anomdimmap}
\end{eqnarray}
We have labelled the variables according to the scheme they correspond to in
order to avoid ambiguity. The mapping indicated via the restriction on the
right hand side means that the expression inside the square brackets is
evaluated in terms of $\MSbar$ variables. However since the left hand side is
the $\RI$ scheme anomalous dimension the $\MSbar$ variables have to be mapped 
to their $\RI$ counterparts. While
\begin{equation}
a_{\RIs} ~=~ a_{\MSbars} 
\label{amap}
\end{equation}
the relation between the gauge parameters is more involved. We have determined
this to the requisite order and note, for brevity, that the $SU(3)$ relation is
\begin{eqnarray}
\left. \frac{}{} \alpha_{\RIs} \right|^{SU(3)} &=&
\left[ 
1
+ \left[
\frac{10}{9} \Nf
- \frac{97}{12}
- \frac{3}{2} \alpha
- \frac{3}{4} \alpha^2
\right] a
\right. \nonumber \\
&& \left. 
+ \left[
27 \zeta_3
- \frac{7143}{32}
- \frac{9}{16} \alpha^3
- \frac{5}{3} \Nf \alpha
- \frac{5}{3} \Nf \alpha^2
+ \frac{4}{3} \zeta_3 \Nf
+ \frac{9}{16} \alpha^4
+ \frac{95}{16} \alpha^2
\right. \right. \nonumber \\
&& \left. \left. ~~~
+ \frac{463}{32} \alpha
+ \frac{2473}{72} \Nf
- 18 \zeta_3 \alpha
\right] a^2
\right. \nonumber \\
&& \left. 
+ \left[
\frac{63225}{64} \zeta_5
- \frac{10221367}{1152}
- \frac{61105}{972} \Nf^2
- \frac{36213}{32} \zeta_3 \alpha
- \frac{21763}{108} \zeta_3 \Nf
- \frac{11417}{288} \Nf \alpha^2
\right. \right. \nonumber \\
&& \left. \left. ~~~
- \frac{8439}{128} \alpha^3
- \frac{3049}{48} \Nf \alpha
- \frac{2320}{9} \zeta_5 \Nf
- \frac{1449}{32} \zeta_3 \alpha^2
- \frac{315}{64} \zeta_5 \alpha^4
- \frac{261}{16} \alpha^4
\right. \right. \nonumber \\
&& \left. \left. ~~~
- \frac{45}{8} \zeta_5 \alpha^3
- \frac{33}{2} \zeta_4 \Nf
- \frac{27}{64} \alpha^6
- \frac{25}{27} \Nf^2 \alpha^2
- \frac{16}{3} \zeta_3 \Nf^2 \alpha
- \frac{11}{4} \zeta_3 \Nf \alpha^2
- \frac{5}{4} \Nf \alpha^3
\right. \right. \nonumber \\
&& \left. \left. ~~~
+ \frac{4}{3} \Nf^2 \alpha
+ \frac{15}{8} \Nf \alpha^4
+ \frac{27}{16} \alpha^5
+ \frac{32}{27} \zeta_3 \Nf^2
+ \frac{81}{8} \zeta_4 \alpha
+ \frac{81}{32} \zeta_4 \alpha^2
+ \frac{117}{32} \zeta_3 \alpha^4
\right. \right. \nonumber \\
&& \left. \left. ~~~
+ \frac{243}{32} \zeta_4
+ \frac{1341}{32} \zeta_3 \alpha^3
+ \frac{3105}{8} \zeta_5 \alpha
+ \frac{3465}{32} \zeta_5 \alpha^2
+ \frac{13941}{8} \zeta_3
+ \frac{30835}{384} \alpha^2
\right. \right. \nonumber \\
&& \left. \left. ~~~
+ \frac{39423}{128} \alpha
+ \frac{4939411}{2592} \Nf
+ 117 \zeta_3 \Nf \alpha
\right] a^3
\right. \nonumber \\
&& \left. 
+ \left[
\frac{230899397}{2048} \zeta_3
+ \frac{246393489}{1024} \zeta_5
- \frac{1174691657}{2304}
- \frac{277127487}{2048} \zeta_7
\right. \right. \nonumber \\
&& \left. \left. ~~~
- \frac{273998929}{31104} \Nf^2
- \frac{70979391}{4096} \zeta_7 \alpha
- \frac{56971325}{2592} \zeta_3 \Nf
- \frac{54117607}{1024} \zeta_3 \alpha
\right. \right. \nonumber \\
&& \left. \left. ~~~
- \frac{18363505}{288} \zeta_5 \Nf
- \frac{13026679}{6912} \Nf \alpha^2
- \frac{12195539}{4608} \Nf \alpha
- \frac{2566431}{2048} \zeta_7 \alpha^2
\right. \right. \nonumber \\
&& \left. \left. ~~~
- \frac{1624581}{512} \zeta_3 \alpha^2
- \frac{1199043}{2048} \alpha^4
- \frac{1100549}{512} \alpha^3
- \frac{992391}{1024} \zeta_5 \alpha^3
- \frac{851175}{1024} \zeta_6 \alpha
\right. \right. \nonumber \\
&& \left. \left. ~~~
- \frac{327213}{512} \zeta_3^2 \alpha
- \frac{305965}{576} \zeta_4 \Nf
- \frac{226449}{512} \zeta_3^2 \alpha^2
- \frac{216815}{96} \zeta_5 \Nf \alpha
\right. \right. \nonumber \\
&& \left. \left. ~~~
- \frac{204525}{32} \zeta_6
- \frac{137781}{1024} \zeta_7 \alpha^4
- \frac{56025}{512} \zeta_3^2 \alpha^3
- \frac{45465}{256} \zeta_5 \alpha^4
- \frac{23247}{1024} \zeta_4 \alpha^3
\right. \right. \nonumber \\
&& \left. \left. ~~~
- \frac{19755}{512} \zeta_5 \alpha^5
- \frac{17631}{512} \zeta_3 \alpha^5
- \frac{13041}{2048} \zeta_4 \alpha^4
- \frac{6071}{96} \Nf \alpha^3
- \frac{5977}{216} \zeta_4 \Nf^2
\right. \right. \nonumber \\
&& \left. \left. ~~~
- \frac{4873}{9} \zeta_3 \Nf^2 \alpha
- \frac{3267}{512} \alpha^5
- \frac{1117}{8} \zeta_3 \Nf \alpha^3
- \frac{880}{27} \zeta_5 \Nf^3
- \frac{567}{256} \alpha^7
- \frac{351}{64} \zeta_3 \alpha^6
\right. \right. \nonumber \\
&& \left. \left. ~~~
- \frac{175}{16} \zeta_5 \Nf \alpha^4
- \frac{128}{3} \zeta_3^2 \Nf^2
- \frac{97}{36} \Nf^2 \alpha^3
- \frac{40}{27} \Nf^3 \alpha
- \frac{15}{8} \Nf \alpha^6
- \frac{7}{2} \zeta_3 \Nf^2 \alpha^2
\right. \right. \nonumber \\
&& \left. \left. ~~~
- \frac{5}{2} \zeta_5 \Nf^2 \alpha^2
+ \frac{9}{4} \zeta_4 \Nf \alpha
+ \frac{9}{4} \zeta_4 \Nf^2 \alpha
+ \frac{9}{4} \zeta_3^2 \Nf \alpha^3
+ \frac{23}{2} \zeta_3 \Nf \alpha^4
+ \frac{25}{12} \Nf^2 \alpha^4
\right. \right. \nonumber \\
&& \left. \left. ~~~
+ \frac{45}{4} \zeta_3^2 \Nf \alpha^2
+ \frac{45}{8} \Nf \alpha^5
+ \frac{81}{256} \alpha^8
+ \frac{115}{2} \zeta_5 \Nf^2 \alpha
+ \frac{153}{4} \zeta_4 \Nf \alpha^2
+ \frac{160}{27} \zeta_3 \Nf^3 \alpha
\right. \right. \nonumber \\
&& \left. \left. ~~~
+ \frac{321}{4} \zeta_3^2 \Nf \alpha
+ \frac{595}{32} \zeta_5 \Nf \alpha^3
+ \frac{945}{128} \zeta_5 \alpha^6
+ \frac{1501}{243} \zeta_3 \Nf^3
+ \frac{1683}{8} \zeta_3^2 \Nf
\right. \right. \nonumber \\
&& \left. \left. ~~~
+ \frac{1765}{128} \Nf \alpha^4
+ \frac{1863}{256} \zeta_3^2 \alpha^5
+ \frac{3069}{128} \alpha^6
+ \frac{4725}{1024} \zeta_6 \alpha^4
+ \frac{10647}{4096} \zeta_7 \alpha^5
\right. \right. \nonumber \\
&& \left. \left. ~~~
+ \frac{16775}{16} \zeta_6 \Nf
+ \frac{19575}{512} \zeta_4 \alpha^2
+ \frac{19737}{512} \zeta_3^2 \alpha^4
+ \frac{22275}{1024} \zeta_6 \alpha^2
+ \frac{24975}{1024} \zeta_6 \alpha^3
\right. \right. \nonumber \\
&& \left. \left. ~~~
+ \frac{26901}{32} \zeta_7 \Nf \alpha
+ \frac{40279}{96} \zeta_5 \Nf \alpha^2
+ \frac{51203}{1296} \Nf^2 \alpha^2
+ \frac{87701}{576} \zeta_3 \Nf \alpha^2
\right. \right. \nonumber \\
&& \left. \left. ~~~
+ \frac{132615}{512} \zeta_7 \alpha^3
+ \frac{197811}{32} \zeta_3^2
+ \frac{285995}{1728} \Nf^2 \alpha
+ \frac{303233}{1944} \zeta_3 \Nf^2
\right. \right. \nonumber \\
&& \left. \left. ~~~
+ \frac{423043}{162} \zeta_5 \Nf^2
+ \frac{488271}{2048} \zeta_3 \alpha^4
+ \frac{636039}{1024} \zeta_4 \alpha
+ \frac{1312185}{512} \alpha^2
\right. \right. \nonumber \\
&& \left. \left. ~~~
+ \frac{2029821}{1024} \zeta_3 \alpha^3
+ \frac{2785621}{288} \zeta_3 \Nf \alpha
+ \frac{2910357}{2048} \zeta_4
+ \frac{4167895}{34992} \Nf^3
\right. \right. \nonumber \\
&& \left. \left. ~~~
+ \frac{4295115}{1024} \zeta_5 \alpha^2
+ \frac{25666081}{864} \zeta_7 \Nf
+ \frac{32185629}{1024} \zeta_5 \alpha
+ \frac{128069033}{18432} \alpha
\right. \right. \nonumber \\
&& \left. \left. ~~~
+ \frac{541249745}{3888} \Nf
+ \zeta_4 \Nf^3
+ 8 \zeta_3 \Nf^2 \alpha^3
\right] a^4
~+~ O(a^5) 
\right] \alpha
\label{alphamap}
\end{eqnarray}
which extends the previous loop order expression given in \cite{13}. In
(\ref{alphamap}) $\zeta_n$ is the Riemann zeta function. The full expression 
for an arbitrary group is available in the data file associated with the arXiv 
version of this article. To effect the mapping indicated on the right side of 
(\ref{anomdimmap}) one has to invert the relation (\ref{alphamap}). The
conversion functions in (\ref{anomdimmap}) are defined by
\begin{equation}
C_\phi(a,\alpha) ~=~ \frac{Z^{\RIs}_\phi}{Z^{\MSbars}_\phi} 
\label{confundef}
\end{equation}
where we use the $\MSbar$ scheme parameters as the argument variables. While 
the renormalization constants are divergent, $C_\phi(a,\alpha)$ is $\epsilon$
finite. This follows by recalling that one has to convert the $\RI$ variables
of the numerator using (\ref{amap}) and (\ref{alphamap}).

\sect{Field anomalous dimensions.}

This section records the field anomalous dimensions at five loops given the
avenue provided by the renormalization group construction (\ref{anomdimmap}).
First, we note that the conversion functions are only needed to four loops
since in (\ref{anomdimmap}) their derivative with respect to either variable is
each multiplied by a renormalization group function. The combination of both
will be $O(a_{\MSbars}^5)$ which is the order we are interested in. In this 
section we will provide results for the $SU(3)$ colour group for brevity and 
therefore note that in the Landau gauge we have
\begin{eqnarray}
\left. \frac{}{} C_A(a,0) \right|^{SU(3)} &=& 1
+ \left[
\frac{97}{12}
- \frac{10}{9} \Nf
\right] a
+ \left[
\frac{100}{81} \Nf^2
+ \frac{83105}{288}
- \frac{11299}{216} \Nf
- \frac{4}{3} \zeta_3 \Nf
- 27 \zeta_3
\right] a^2
\nonumber \\
&&
+ \left[
\frac{16}{9} \zeta_3 \Nf^2
- \frac{8228977}{2592} \Nf
- \frac{63225}{64} \zeta_5
- \frac{17433}{8} \zeta_3
- \frac{1000}{729} \Nf^3
- \frac{243}{32} \zeta_4
\right. \nonumber \\
&& \left. ~~~
+ \frac{33}{2} \zeta_4 \Nf
+ \frac{2320}{9} \zeta_5 \Nf
+ \frac{25915}{108} \zeta_3 \Nf
+ \frac{164395}{972} \Nf^2
+ \frac{44961125}{3456}
\right] a^3
\nonumber \\
&&
+ \left[
\frac{62302764631}{82944}
- \frac{1758762815}{7776} \Nf
- \frac{324121925}{2048} \zeta_3
- \frac{262747689}{1024} \zeta_5
\right. \nonumber \\
&& \left. ~~~
- \frac{25666081}{864} \zeta_7 \Nf
- \frac{15059695}{34992} \Nf^3
- \frac{3161781}{2048} \zeta_4
- \frac{515843}{162} \zeta_5 \Nf^2
\right. \nonumber \\
&& \left. ~~~
- \frac{362555}{648} \zeta_3 \Nf^2
- \frac{174483}{32} \zeta_3^2
- \frac{16775}{16} \zeta_6 \Nf
- \frac{1943}{216} \zeta_4 \Nf^2
\right. \nonumber \\
&& \left. ~~~
- \frac{1107}{8} \zeta_3^2 \Nf
- \frac{229}{27} \zeta_3 \Nf^3
+ \frac{400}{9} \zeta_3^2 \Nf^2
+ \frac{880}{27} \zeta_5 \Nf^3
\right. \nonumber \\
&& \left. ~~~
+ \frac{10000}{6561} \Nf^4
+ \frac{204525}{32} \zeta_6
+ \frac{469333}{576} \zeta_4 \Nf
+ \frac{27270475}{864} \zeta_3 \Nf
\right. \nonumber \\
&& \left. ~~~
+ \frac{60587905}{864} \zeta_5 \Nf
+ \frac{190787741}{10368} \Nf^2
+ \frac{277127487}{2048} \zeta_7
- \zeta_4 \Nf^3
\right] a^4
\nonumber \\
&& +~ O(a^5)
\end{eqnarray}
\begin{eqnarray}
\left. \frac{}{} C_c(a,0) \right|^{SU(3)} &=& 1
+ 3 a
+ \left[
\frac{5829}{64}
- \frac{135}{16} \zeta_3
- \frac{95}{16} \Nf
\right] a^2
\nonumber \\
&&
+ \left[
\frac{5161}{648} \Nf^2
- \frac{198001}{432} \Nf
- \frac{40449}{64} \zeta_3
- \frac{1755}{32} \zeta_5
- \frac{33}{4} \zeta_4 \Nf
+ \frac{2}{3} \zeta_3 \Nf^2
\right. \nonumber \\
&& \left. ~~~
+ \frac{59}{2} \zeta_3 \Nf
+ \frac{243}{64} \zeta_4
+ \frac{1082353}{288}
\right] a^3
\nonumber \\
&&
+ \left[
\frac{15567976783}{73728}
- \frac{151911987}{4096} \zeta_3
- \frac{79190001}{2048} \zeta_5
- \frac{39621021}{1024} \Nf
- \frac{204525}{64} \zeta_6
\right. \nonumber \\
&& \left. ~~~
- \frac{150979}{7776} \Nf^3
- \frac{52025}{128} \zeta_4 \Nf
- \frac{11907}{16} \zeta_7 \Nf
- \frac{699}{8} \zeta_5 \Nf^2
- \frac{425}{48} \zeta_3 \Nf^2
\right. \nonumber \\
&& \left. ~~~
- \frac{5}{18} \zeta_3 \Nf^3
+ \frac{1}{2} \zeta_4 \Nf^3
+ \frac{165}{16} \zeta_4 \Nf^2
+ \frac{337}{4} \zeta_3^2 \Nf
+ \frac{16775}{32} \zeta_6 \Nf
\right. \nonumber \\
&& \left. ~~~
+ \frac{250085}{64} \zeta_5 \Nf
+ \frac{1149471}{512} \zeta_3^2
+ \frac{2625583}{768} \zeta_3 \Nf
+ \frac{5819877}{4096} \zeta_4
\right. \nonumber \\
&& \left. ~~~
+ \frac{8093153}{4608} \Nf^2
+ \frac{100880073}{8192} \zeta_7
\right] a^4 ~+~ O(a^5)
\end{eqnarray}
and
\begin{eqnarray}
\left. \frac{}{} C_\psi(a,0) \right|^{SU(3)} &=& 1
+ \left[
12 \zeta_3
- \frac{359}{9}
+ \frac{7}{3} \Nf
\right] a^2
\nonumber \\
&&
+ \left[
\frac{24722}{81} \Nf
- \frac{439543}{162}
- \frac{1570}{243} \Nf^2
- \frac{1165}{3} \zeta_5
- \frac{440}{9} \zeta_3 \Nf
\right. \nonumber \\
&& \left. ~~~
+ \frac{79}{4} \zeta_4
+ \frac{8009}{6} \zeta_3
\right] a^3
\nonumber \\
&&
+ \left[
\frac{21391}{1458} \Nf^3
- \frac{356864009}{5184} \zeta_5
- \frac{146722043}{864}
- \frac{29889697}{5184} \zeta_3^2
- \frac{1294381}{108} \zeta_3 \Nf
\right. \nonumber \\
&& \left. ~~~
- \frac{1276817}{972} \Nf^2
- \frac{440}{9} \zeta_5 \Nf^2
- \frac{20}{3} \zeta_4 \Nf^2
+ \frac{8}{27} \zeta_3 \Nf^3
+ \frac{100}{3} \zeta_6 \Nf
+ \frac{2291}{72} \zeta_4 \Nf
\right. \nonumber \\
&& \left. ~~~
+ \frac{5704}{27} \zeta_3 \Nf^2
+ \frac{565939}{864} \zeta_4
+ \frac{1673051}{324} \zeta_5 \Nf
+ \frac{3807625}{10368} \zeta_6
+ \frac{6747755}{288} \zeta_7
\right. \nonumber \\
&& \left. ~~~
+ \frac{55476671}{1944} \Nf
+ \frac{317781451}{2592} \zeta_3
- 1029 \zeta_7 \Nf
- 24 \zeta_3^2 \Nf
\right] a^4 ~+~ O(a^5)
\end{eqnarray}
with the expressions for non-zero $\alpha$ and an arbitrary Lie group available
in the article's data file. With these it is therefore a straightforward 
exercise to deduce
\begin{eqnarray}
\left. \gamma_A^{\RIs}(a,0) \right|^{SU(3)} &=&
\left[
\frac{2}{3} \Nf
- \frac{13}{2}
\right] a
+ \left[
\frac{250}{9} \Nf
- \frac{3727}{24}
- \frac{20}{27} \Nf^2
\right] a^2
\nonumber \\
&&
+ \left[
\frac{200}{243} \Nf^3
+ \frac{5210}{3} \Nf
+ \frac{9747}{16} \zeta_3
- \frac{2127823}{288}
- \frac{1681}{18} \Nf^2
- \frac{119}{3} \zeta_3 \Nf
\right. \nonumber \\
&& \left. ~~~
- \frac{16}{9} \zeta_3 \Nf^2
\right] a^3
\nonumber \\
&&
+ \left[
\frac{373823}{1458} \Nf^3
- \frac{3011547563}{6912}
- \frac{6816713}{648} \Nf^2
- \frac{2897113}{216} \zeta_3 \Nf
\right. \nonumber \\
&& \left. ~~~
- \frac{845275}{96} \zeta_5 \Nf
- \frac{2000}{2187} \Nf^4
+ \frac{88}{27} \zeta_3 \Nf^3
+ \frac{4640}{9} \zeta_5 \Nf^2
+ \frac{60427}{162} \zeta_3 \Nf^2
\right. \nonumber \\
&& \left. ~~~
+ \frac{1431945}{64} \zeta_5
+ \frac{18987543}{256} \zeta_3
+ \frac{221198219}{1728} \Nf
\right] a^4
\nonumber \\
&&
+ \left[
\frac{1791101885}{34992} \Nf^3
- \frac{5296290721381}{165888}
- \frac{88396975485}{16384} \zeta_7
\right. \nonumber \\
&& \left. ~~~
- \frac{28725816895}{23328} \Nf^2
- \frac{4380999739}{2592} \zeta_3 \Nf
- \frac{4144018255}{1152} \zeta_5 \Nf
\right. \nonumber \\
&& \left. ~~~
+ \frac{11001160807}{6912} \zeta_7 \Nf
+ \frac{13433378467}{2048} \zeta_3
+ \frac{42793732125}{4096} \zeta_5
\right. \nonumber \\
&& \left. ~~~
+ \frac{51527836699}{4608} \Nf
- \frac{25666081}{324} \zeta_7 \Nf^2
- \frac{18057505}{26244} \Nf^4
- \frac{942553}{64} \zeta_3^2 \Nf
\right. \nonumber \\
&& \left. ~~~
- \frac{761560}{81} \zeta_5 \Nf^3
- \frac{34749}{4} \zeta_4
- \frac{16744}{81} \zeta_3 \Nf^3
- \frac{13771}{54} \zeta_4 \Nf^2
\right. \nonumber \\
&& \left. ~~~
- \frac{13456}{9} \zeta_3^2 \Nf^2
- \frac{4120}{243} \zeta_3 \Nf^4
+ \frac{1618}{81} \zeta_4 \Nf^3
+ \frac{3136}{27} \zeta_3^2 \Nf^3
+ \frac{7040}{81} \zeta_5 \Nf^4
\right. \nonumber \\
&& \left. ~~~
+ \frac{20000}{19683} \Nf^5
+ \frac{53227}{36} \zeta_4 \Nf
+ \frac{137558185}{432} \zeta_5 \Nf^2
+ \frac{231090011}{2592} \zeta_3 \Nf^2
\right. \nonumber \\
&& \left. ~~~
+ \frac{589719519}{2048} \zeta_3^2
\right] a^5 ~+~ O(a^6)
\end{eqnarray}
\begin{eqnarray}
\left. \gamma_c^{\RIs}(a,0) \right|^{SU(3)} &=&
-~ \frac{9}{4} a
+ \left[
\frac{13}{4} \Nf
- \frac{813}{16}
\right] a^2
\nonumber \\
&&
+ \left[
\frac{21}{4} \zeta_3 \Nf
+ \frac{14909}{48} \Nf
+ \frac{5697}{32} \zeta_3
- \frac{157303}{64}
- \frac{125}{18} \Nf^2
\right] a^3
\nonumber \\
&&
+ \left[
\frac{2705}{162} \Nf^3
- \frac{219384137}{1536}
- \frac{288155}{216} \Nf^2
- \frac{132749}{96} \zeta_3 \Nf
- \frac{15175}{16} \zeta_5 \Nf
\right. \nonumber \\
&& \left. ~~~
+ \frac{19}{2} \zeta_3 \Nf^2
+ \frac{221535}{32} \zeta_5
+ \frac{9207729}{512} \zeta_3
+ \frac{30925009}{1152} \Nf
\right] a^4
\nonumber \\
&&
+ \left[
\frac{44742833}{7776} \Nf^3
- \frac{373954139515}{36864}
- \frac{26906792787}{32768} \zeta_7
+ \frac{6421603437}{4096} \zeta_3
\right. \nonumber \\
&& \left. ~~~
+ \frac{18444521475}{8192} \zeta_5
+ \frac{69770872099}{27648} \Nf
- \frac{463728159}{4096} \zeta_3^2
- \frac{280840685}{768} \zeta_5 \Nf
\right. \nonumber \\
&& \left. ~~~
- \frac{90397253}{384} \zeta_3 \Nf
- \frac{28603559}{144} \Nf^2
- \frac{37306}{729} \Nf^4
- \frac{24057}{8} \zeta_4
- \frac{3969}{2} \zeta_7 \Nf^2
\right. \nonumber \\
&& \left. ~~~
- \frac{809}{12} \zeta_4 \Nf^2
- \frac{437}{2} \zeta_3^2 \Nf^2
- \frac{415}{3} \zeta_5 \Nf^3
- \frac{172}{3} \zeta_3 \Nf^3
+ \frac{1627}{8} \zeta_4 \Nf
\right. \nonumber \\
&& \left. ~~~
+ \frac{498713}{64} \zeta_3 \Nf^2
+ \frac{720477}{128} \zeta_3^2 \Nf
+ \frac{4411505}{288} \zeta_5 \Nf^2
+ \frac{318680873}{3072} \zeta_7 \Nf
\right] a^5
\nonumber \\
&& +~ O(a^6)
\end{eqnarray}
and
\begin{eqnarray}
\left. \gamma_\psi^{\RIs}(a,0) \right|^{SU(3)} &=&
\left[
\frac{67}{3}
- \frac{4}{3} \Nf
\right] a^2
+ \left[
\frac{104}{27} \Nf^2
+ \frac{52321}{36}
+ 16 \zeta_3 \Nf
- \frac{4472}{27} \Nf
- \frac{607}{2} \zeta_3
\right] a^3
\nonumber \\
&&
+ \left[
\frac{75355}{81} \Nf^2
- \frac{15631129}{324} \zeta_3
- \frac{1537880}{81} \Nf
- \frac{1000}{81} \Nf^3
+ \frac{121558}{27} \zeta_3 \Nf
\right. \nonumber \\
&& \left. ~~~
+ \frac{8966278}{81}
+ \frac{15846715}{1296} \zeta_5
- 830 \zeta_5 \Nf
- 80 \zeta_3 \Nf^2
\right] a^4
\nonumber \\
&&
+ \left[
\frac{86560}{2187} \Nf^4
- \frac{61222404563}{10368} \zeta_3
- \frac{26588447977}{27648} \zeta_7
- \frac{23362550497}{11664} \Nf
\right. \nonumber \\
&& \left. ~~~
+ \frac{2727697601}{10368} \zeta_3^2
+ \frac{2947769225}{972} \zeta_5
+ \frac{10680056713}{11664} \zeta_3 \Nf
\right. \nonumber \\
&& \left. ~~~
+ \frac{136110640603}{15552}
- \frac{1254469405}{2916} \zeta_5 \Nf
- \frac{9978134}{243} \zeta_3 \Nf^2
\right. \nonumber \\
&& \left. ~~~
- \frac{9708268}{2187} \Nf^3
- \frac{645017}{54} \zeta_3^2 \Nf
- \frac{6811}{2} \zeta_7 \Nf^2
- \frac{3520}{27} \zeta_5 \Nf^3
- \frac{704}{9} \zeta_3^2 \Nf^2
\right. \nonumber \\
&& \left. ~~~
+ \frac{43264}{81} \zeta_3 \Nf^3
+ \frac{3815110}{243} \zeta_5 \Nf^2
+ \frac{4429579}{36} \zeta_7 \Nf
+ \frac{222450779}{1458} \Nf^2
\right] a^5
\nonumber \\
&& +~ O(a^6) 
\end{eqnarray}
at five loops. We have checked that the previous three loop results of 
\cite{11,12,13} are reproduced since the four loop renormalization constants 
were derived from the {\sc Forcer} data provided in \cite{40}. These relations 
now together with the five loop $\RI$ $\beta$-function complete the derivation 
of the $\RI$ scheme renormalization group functions to that order. More 
importantly we have determined the $\RI$ renormalization constant for the quark
itself. This is one of our main goals since it is central to the 
renormalization of quark bilinear operators in the $\RI$ scheme that are of 
importance in lattice matching analyses.

\sect{Quark mass operator.}

We now turn to the determination of the Green's functions with zero momentum
operator insertions which also extend the results provided in \cite{13}. Unlike
the derivation of the five loop anomalous dimensions of the fields in the $\RI$
scheme of the previous section no expression is available for the Green's 
function as a function of the bare parameters that is measured on the lattice 
involving the quark mass operator. Therefore we have computed the four loop 
corrections explicitly for the present analysis. First, to be specific the 
Green's function of the quark mass operator we will consider is 
\begin{equation}
G_{\bar{\psi}\psi}(p) ~=~ \langle \psi(p) ~ [ \bar{\psi} \psi ](0) ~ 
\bar{\psi}(-p) \rangle ~.
\label{gfmass}
\end{equation}
As $G_{\bar{\psi}\psi}(p)$ involves a Lorentz scalar operator it has only one
form factor which can be accessed by taking the spinor trace, normalized by the
trace over the unit matrix. This produces Feynman integrals that are Lorentz 
scalars which means we can apply the four loop automatic Feynman graph
computation programme {\sc Forcer}, \cite{36,37}, to it. Previously, the three
loop {\sc Mincer} programme was used in \cite{13}. Here we have computed the
$5728$ four loop graphs contributing to (\ref{gfmass}) as well as the $1$, $13$
and $244$ diagrams respectively at one, two and three loops. The lower loop 
graphs were calculated with the {\sc Forcer} algorithm both for consistency and
as a check on previous work. In addition to using {\sc Forcer} to evaluate the 
large number of graphs, that are generated electronically using {\sc Qgraf} 
\cite{44}, we used the {\sc Form} {\tt color.h} module \cite{41} since it 
automatically determines the colour group factor associated with each graph 
prior to integration. The module is based on \cite{45}. Consequently we have 
evaluated 
\begin{equation}
\Sigma^{(1)}_{\bar{\psi}\psi}(p) ~=~
\frac{1}{4} \mbox{tr} \left[ G_{\bar{\psi}\psi}(p) \right] 
\end{equation}
where $\mbox{tr}$ is the Lorentz spinor trace and found
\begin{eqnarray}
\left. \Sigma^{(1) ~ \MSbars}_{\bar{\psi}\psi}(p) \right|_{\alpha=0}
&=& 1 ~+~ 4 C_F a
\nonumber \\
&&
+~ \left[  
\frac{1531}{24} C_F C_A
+ 13 C_F^2
- \frac{52}{3} \Nf T_F C_F
- 21 \zeta_3 C_F C_A
+ 12 \zeta_3 C_F^2
\right] a^2
\nonumber \\
&&
+~ \left[  
\frac{4315565}{3888} C_F C_A^2
+ \frac{3005}{18} C_F^2 C_A
+ \frac{916}{3} C_F^3
- \frac{131048}{243} \Nf T_F C_F C_A
\right. \nonumber \\
&& \left. ~~~~~
- \frac{4699}{18} \Nf T_F C_F^2
+ \frac{12224}{243} \Nf^2 T_F^2 C_F
+ \frac{405}{4} \zeta_5 C_F C_A^2
- 40 \zeta_5 C_F^2 C_A
\right. \nonumber \\
&& \left. ~~~~~
- 120 \zeta_5 C_F^3
- \frac{69}{16} \zeta_4 C_F C_A^2
+ 6 \zeta_4 C_F^2 C_A
- 24 \zeta_4 \Nf T_F C_F C_A
\right. \nonumber \\
&& \left. ~~~~~
+ 24 \zeta_4 \Nf T_F C_F^2
- \frac{20305}{36} \zeta_3 C_F C_A^2
+ \frac{626}{3} \zeta_3 C_F^2 C_A
+ 38 \zeta_3 C_F^3
\right. \nonumber \\
&& \left. ~~~~~
+ \frac{772}{9} \zeta_3 \Nf T_F C_F C_A
+ \frac{176}{3} \zeta_3 \Nf T_F C_F^2
+ \frac{32}{9} \zeta_3 \Nf^2 T_F^2 C_F
\right] a^3
\nonumber \\
&&
+~ \left[  
\frac{2200233199}{93312} C_F C_A^3
- \frac{759}{2} \frac{d_F^{abcd} d_A^{abcd}}{\NF}
- \frac{41391917}{31104} C_F^2 C_A^2
\right. \nonumber \\
&& \left. ~~~~~
+ \frac{1095131}{72} C_F^3 C_A
+ \frac{2033}{12} C_F^4
+ 808 \Nf \frac{d_F^{abcd} d_F^{abcd}}{\NF}
\right. \nonumber \\
&& \left. ~~~~~
- \frac{66598003}{3888} \Nf T_F C_F C_A^2
- \frac{33045443}{3888} \Nf T_F C_F^2 C_A
- \frac{305695}{72} \Nf T_F C_F^3
\right. \nonumber \\
&& \left. ~~~~~
+ \frac{7064407}{1944} \Nf^2 T_F^2 C_F C_A
+ \frac{1017025}{486} \Nf^2 T_F^2 C_F^2
- \frac{161152}{729} \Nf^3 T_F^3 C_F
\right. \nonumber \\
&& \left. ~~~~~
+ \frac{43295}{16} \zeta_7 \frac{d_F^{abcd} d_A^{abcd}}{\NF}
- \frac{2227141}{768} \zeta_7 C_F C_A^3
+ \frac{365813}{32} \zeta_7 C_F^2 C_A^2
\right. \nonumber \\
&& \left. ~~~~~
- \frac{43351}{2} \zeta_7 C_F^3 C_A
+ 15078 \zeta_7 C_F^4
+ 1764 \zeta_7 \Nf \frac{d_F^{abcd} d_F^{abcd}}{\NF}
\right. \nonumber \\
&& \left. ~~~~~
- \frac{2499}{4} \zeta_7 \Nf T_F C_F C_A^2
+ 1764 \zeta_7 \Nf T_F C_F^2 C_A
- \frac{4275}{64} \zeta_6 \frac{d_F^{abcd} d_A^{abcd}}{\NF}
\right. \nonumber \\
&& \left. ~~~~~
- \frac{91125}{256} \zeta_6 C_F C_A^3
+ \frac{6125}{8} \zeta_6 C_F^2 C_A^2
- 900 \zeta_6 C_F^3 C_A
+ 400 \zeta_6 C_F^4
\right. \nonumber \\
&& \left. ~~~~~
+ 200 \zeta_6 \Nf T_F C_F C_A^2
+ 150 \zeta_6 \Nf T_F C_F^2 C_A
- 300 \zeta_6 \Nf T_F C_F^3
\right. \nonumber \\
&& \left. ~~~~~
+ \frac{7075}{32} \zeta_5 \frac{d_F^{abcd} d_A^{abcd}}{\NF}
+ \frac{2673791}{1152} \zeta_5 C_F C_A^3
+ 203 \zeta_5 C_F^2 C_A^2
- 2225 \zeta_5 C_F^3 C_A
\right. \nonumber \\
&& \left. ~~~~~
- 5950 \zeta_5 C_F^4
- 1440 \zeta_5 \Nf \frac{d_F^{abcd} d_F^{abcd}}{\NF}
- \frac{5933}{72} \zeta_5 \Nf T_F C_F C_A^2
\right. \nonumber \\
&& \left. ~~~~~
- 170 \zeta_5 \Nf T_F C_F^2 C_A
+ 360 \zeta_5 \Nf T_F C_F^3
- \frac{1688}{9} \zeta_5 \Nf^2 T_F^2 C_F C_A
\right. \nonumber \\
&& \left. ~~~~~
+ 112 \zeta_5 \Nf^2 T_F^2 C_F^2
+ \frac{9099}{64} \zeta_4 \frac{d_F^{abcd} d_A^{abcd}}{\NF}
+ \frac{34871}{1536} \zeta_4 C_F C_A^3
\right. \nonumber \\
&& \left. ~~~~~
- \frac{2451}{16} \zeta_4 C_F^2 C_A^2
+ \frac{921}{2} \zeta_4 C_F^3 C_A
- 276 \zeta_4 C_F^4
- 180 \zeta_4 \Nf \frac{d_F^{abcd} d_F^{abcd}}{\NF}
\right. \nonumber \\
&& \left. ~~~~~
- \frac{5377}{16} \zeta_4 \Nf T_F C_F C_A^2
+ 42 \zeta_4 \Nf T_F C_F^2 C_A
+ 279 \zeta_4 \Nf T_F C_F^3
\right. \nonumber \\
&& \left. ~~~~~
+ 72 \zeta_4 \Nf^2 T_F^2 C_F C_A
- 72 \zeta_4 \Nf^2 T_F^2 C_F^2
+ \frac{16}{3} \zeta_4 \Nf^3 T_F^3 C_F
\right. \nonumber \\
&& \left. ~~~~~
+ \frac{21183}{16} \zeta_3 \frac{d_F^{abcd} d_A^{abcd}}{\NF}
- \frac{24776933}{1728} \zeta_3 C_F C_A^3
- \frac{1457}{8} \zeta_3 C_F^2 C_A^2
\right. \nonumber \\
&& \left. ~~~~~
+ 9049 \zeta_3 C_F^3 C_A
- 4652 \zeta_3 C_F^4
- 552 \zeta_3 \Nf \frac{d_F^{abcd} d_F^{abcd}}{\NF}
\right. \nonumber \\
&& \left. ~~~~~
+ \frac{432649}{72} \zeta_3 \Nf T_F C_F C_A^2
+ \frac{3155}{3} \zeta_3 \Nf T_F C_F^2 C_A
+ 1226 \zeta_3 \Nf T_F C_F^3
\right. \nonumber \\
&& \left. ~~~~~
- \frac{556}{3} \zeta_3 \Nf^2 T_F^2 C_F C_A
- \frac{2920}{3} \zeta_3 \Nf^2 T_F^2 C_F^2
- \frac{128}{27} \zeta_3 \Nf^3 T_F^3 C_F
\right. \nonumber \\
&& \left. ~~~~~
- \frac{74031}{32} \zeta_3^2 \frac{d_F^{abcd} d_A^{abcd}}{\NF}
+ \frac{101263}{64} \zeta_3^2 C_F C_A^3
- \frac{11359}{4} \zeta_3^2 C_F^2 C_A^2
\right. \nonumber \\
&& \left. ~~~~~
+ 3320 \zeta_3^2 C_F^3 C_A
- 1472 \zeta_3^2 C_F^4
- 192 \zeta_3^2 \Nf \frac{d_F^{abcd} d_F^{abcd}}{\NF}
\right. \nonumber \\
&& \left. ~~~~~
+ 384 \zeta_3^2 \Nf T_F C_F C_A^2
- 812 \zeta_3^2 \Nf T_F C_F^2 C_A
+ 200 \zeta_3^2 \Nf T_F C_F^3
\right] a^4
\nonumber \\
&& +~ O(a^5)
\label{gfmassgen}
\end{eqnarray} 
in the Landau gauge in the $\MSbar$ scheme for an arbitrary colour group where
$\NF$ is the dimension of the fundamental representation. The result for an 
arbitrary gauge is included in the associated data file. Aside from the usual 
colour factors $T_F$, $C_F$ and $C_A$ the rank $4$ fully symmetric tensor 
$d_R^{abcd}$ appears for both the fundamental $F$ and adjoint $A$ 
representations. Their properties are given in \cite{45} and they are defined 
by
\begin{equation}
d_R^{abcd} ~=~ \frac{1}{6} \mbox{Tr} \left( T^a T^{(b} T^c T^{d)}
\right)
\end{equation}
for the group generators $T^a$ in representation $R$. The trace $\mbox{Tr}$ is
over the colour spinor indices. In determining (\ref{gfmassgen}) we have 
verified the $\MSbar$ four loop quark mass anomalous dimension found in
\cite{31,32,46} is reproduced as a check on our procedure. For practical 
applications to lattice matching we note that (\ref{gfmassgen}) numerically 
evaluates to
\begin{eqnarray}
\left. \Sigma^{(1) ~ \MSbars}_{\bar{\psi}\psi}(p) 
\right|_{\alpha=0}^{SU(3)} &=&
1 + 5.333333 a + [ 202.948878 - 11.555556 \Nf ] a^2 
\nonumber \\
&&
+~ [ 18.192836 \Nf^2 - 1070.591116 \Nf + 8966.208391 ] a^3 
\nonumber \\
&&
+~ [ 439203.615244 - 81491.908465 \Nf + 3721.663435 \Nf^2 
- 36.830872 \Nf^3 ] a^4 
\nonumber \\
&&
+~ O(a^5)
\label{gfmasssu3num}
\end{eqnarray} 
for $SU(3)$. As the four loop correction is important in assisting error
analysis for lattice matching it is worthwhile estimating the effect of the new
term. If we take the value of the strong coupling $\alpha_s$~$=$~$g^2/(4\pi)$
to be $\alpha_s$~$=$~$0.12$ when $\Nf$~$=$~$3$ then at successive loop orders 
(\ref{gfmasssu3num}) gives $1.05092958$, $1.06627508$, $1.07142857$ and 
$1.07331808$ respectively. The difference between the two and three loop values
is around $0.5\%$ but between three and four loops this drops to around 
$0.2\%$. This rough comparison suggests that the four loop expression could 
refine matching errors for quark mass measurements.

Having established one of our main results we turn now to the other scheme of
interest which is $\RI$. We recall the condition to renormalize the quark mass
operator in that scheme is
\begin{equation}
\frac{1}{4}
\left. \lim_{\epsilon \rightarrow 0} \mbox{tr} \left[
Z^{\mbox{\footnotesize{RI$^\prime$}}}_{\bar{\psi}\psi}
Z^{\mbox{\footnotesize{RI$^\prime$}}}_\psi \langle \psi(p) 
[ \bar{\psi} \psi ](0)  \bar{\psi}(-p) \rangle \right]
\right|_{p^2 \, = \, \mu^2} ~=~ 1
\end{equation}
which runs parallel to that for the quark wave function renormalization.
Applying this to the value we obtained for the Green's function where the 
parameters are bare allows us to find 
$Z^{\mbox{\footnotesize{RI$^\prime$}}}_{\bar{\psi}\psi}$ at four loops. Using
the renormalization group procedure of the previous section means we can deduce
the four loop contribution to the quark mass conversion that was given in 
\cite{11,12,13} at lower order. For brevity we note the expression in the 
Landau gauge for $SU(3)$ is 
\begin{eqnarray}
\left. \frac{}{} C_m(a,0) \right|^{SU(3)} &=& 1
- \frac{16}{3} a
+ \left[
\frac{83}{9} \Nf
+ \frac{152}{3} \zeta_3
- \frac{3779}{18}
\right] a^2
\nonumber \\
&&
+ \left[
\frac{217390}{243} \Nf
- \frac{3115807}{324}
- \frac{7514}{729} \Nf^2
- \frac{4720}{27} \zeta_3 \Nf
- \frac{2960}{9} \zeta_5
- \frac{32}{27} \zeta_3 \Nf^2
\right. \nonumber \\
&& \left. ~~~
+ \frac{80}{3} \zeta_4 \Nf
+ \frac{195809}{54} \zeta_3
\right] a^3
\nonumber \\
&&
+ \left[
\frac{96979}{4374} \Nf^3
- \frac{744609145}{1296}
- \frac{52383125}{17496} \Nf^2
- \frac{3837631}{1728} \zeta_7
- \frac{2017309}{81} \zeta_3 \Nf
\right. \nonumber \\
&& \left. ~~~
- \frac{843077}{54} \zeta_3^2
- \frac{359855}{81} \zeta_5 \Nf
- \frac{16960}{9} \zeta_4
- \frac{11500}{9} \zeta_6 \Nf
- \frac{8776}{27} \zeta_3^2 \Nf
\right. \nonumber \\
&& \left. ~~~
- \frac{343}{2} \zeta_7 \Nf
- \frac{100}{3} \zeta_4 \Nf^2
- \frac{8}{9} \zeta_4 \Nf^3
+ \frac{40}{81} \zeta_3 \Nf^3
+ \frac{560}{3} \zeta_5 \Nf^2
+ \frac{11542}{9} \zeta_4 \Nf
\right. \nonumber \\
&& \left. ~~~
+ \frac{33964}{81} \zeta_3 \Nf^2
+ \frac{9369745}{432} \zeta_5
+ \frac{86284171}{324} \zeta_3
+ \frac{247516535}{2916} \Nf
+ 5500 \zeta_6
\right] a^4
\nonumber \\
&& +~ O(a^5)
\end{eqnarray}
where the full colour group and gauge dependence for this and all the other
results in this section are provided in the associated data file. Similar to
the last section knowing the $\MSbar$ five loop quark mass anomalous dimension,
\cite{35}, means we can deduce the five loop $\RI$ scheme counterpart which is 
\begin{eqnarray}
\left. \gamma_m^{\RIs}(a,0) \right|^{SU(3)} &=&
-~ 4 a
+ \left[
\frac{52}{9} \Nf
- 126
\right] a^2
\nonumber \\
&&
+ \left[
\frac{17588}{27} \Nf
+ \frac{3344}{3} \zeta_3
- \frac{20174}{3}
- \frac{856}{81} \Nf^2
- \frac{128}{9} \zeta_3 \Nf
\right] a^3
\nonumber \\
&&
+ \left[
\frac{16024}{729} \Nf^3
- \frac{141825253}{324}
- \frac{611152}{243} \Nf^2
- \frac{298241}{27} \zeta_3 \Nf
- \frac{6160}{3} \zeta_5
\right. \nonumber \\
&& \left. ~~~
- \frac{4160}{3} \zeta_5 \Nf
+ \frac{5984}{27} \zeta_3 \Nf^2
+ \frac{3519059}{54} \Nf
+ \frac{7230017}{54} \zeta_3
\right] a^4
\nonumber \\
&&
+  \left[
\frac{3457170034}{243} \zeta_3
- \frac{32529575441}{972}
- \frac{492112429}{243} \zeta_3 \Nf
- \frac{314416490}{729} \Nf^2
\right. \nonumber \\
&& \left. ~~~
- \frac{220508981}{432} \zeta_7
- \frac{214516115}{486} \zeta_5 \Nf
- \frac{22686158}{27} \zeta_3^2
- \frac{380896}{6561} \Nf^4
\right. \nonumber \\
&& \left. ~~~
- \frac{289936}{243} \zeta_3 \Nf^3
- \frac{68992}{81} \zeta_3^2 \Nf^2
- \frac{8320}{27} \zeta_5 \Nf^3
- \frac{3236}{27} \zeta_4 \Nf^2
\right. \nonumber \\
&& \left. ~~~
+ \frac{1372}{3} \zeta_7 \Nf^2
+ \frac{3254}{9} \zeta_4 \Nf
+ \frac{3172244}{81} \zeta_3^2 \Nf
+ \frac{5666800}{243} \zeta_5 \Nf^2
\right. \nonumber \\
&& \left. ~~~
+ \frac{14071511}{1458} \Nf^3
+ \frac{21021016}{243} \zeta_3 \Nf^2
+ \frac{42627823}{648} \zeta_7 \Nf
+ \frac{1646302180}{243} \Nf
\right. \nonumber \\
&& \left. ~~~
+ \frac{1681793075}{972} \zeta_5
- 5346 \zeta_4
\right] a^5 ~+~ O(a^6)
\end{eqnarray}
for the same configuration. Finally we note
\begin{eqnarray}
\Sigma^{(1) ~ \RIs}_{\bar{\psi}\psi}(p) ~=~ 1
\end{eqnarray}
follows trivially from the definition of the $\RI$ scheme.

\sect{Vector current.}

We now turn to the other operator of interest which is the flavour non-singlet 
vector current $\bar{\psi} \gamma^\mu \psi$ and, in a similar procedure to that
of the previous section, determine the analogous Green's function to 
(\ref{gfmass}) which is
\begin{equation}
G^\mu_{\bar{\psi} \gamma^\mu \psi}(p) ~=~ \langle \psi(p) ~ [ \bar{\psi}
\gamma^\mu \psi ] (0) ~ \bar{\psi}(-p) \rangle ~=~ \Sigma^{(1)}_{\bar{\psi}
\gamma^\mu \psi}(p) \gamma^\mu ~+~ \Sigma^{(2)}_{\bar{\psi} \gamma^\mu \psi}(p)
\frac{p^\mu \pslash}{p^2} ~.
\label{gfvec}
\end{equation}
As the operator is clearly a Lorentz vector its zero momentum insertion in the
quark $2$-point function has to decompose into a basis of vectors depending on 
the external momenta and $\gamma^\mu$ by Lorentz symmetry. This is reflected in
the two form factors $\Sigma^{(i)}_{\bar{\psi} \gamma^\mu \psi}(p)$ of
(\ref{gfvec}) which are Lorentz scalars. To evaluate them to four loops using 
{\sc Forcer}, which can only be applied to scalar Feynman integrals, we project
out the form factors via the relations
\begin{eqnarray}
\Sigma^{(1)}_{\bar{\psi} \gamma^\mu \psi}(p) &=& \frac{1}{4(d-1)} \left[
\mbox{tr} \left( \gamma_\mu G^\mu_{\bar{\psi} \gamma^\mu \psi}(p) \right) ~-~
\mbox{tr} \left( \frac{p_\mu \pslash}{p^2} G^\mu_{\bar{\psi} \gamma^\mu
\psi}(p) \right) \right] \nonumber \\
\Sigma^{(2)}_{\bar{\psi} \gamma^\mu \psi}(p) &=& -~ \frac{1}{4(d-1)} \left[
\mbox{tr} \left( \gamma_\mu G^\mu_{\bar{\psi} \gamma^\mu \psi}(p) \right) ~-~
d \, \mbox{tr} \left( \frac{p_\mu \pslash}{p^2} G^\mu_{\bar{\psi} \gamma^\mu
\psi}(p) \right) \right] 
\end{eqnarray}
in the same way as \cite{13}. One property of the non-singlet vector current is
that since it is a physical operator that relates to charge conservation its
anomalous dimension is zero. In other words the renormalization constant of the
current itself is unity in {\em all} renormalization schemes meaning
$\Sigma^{(i)}_{\bar{\psi} \gamma^\mu \psi}(p)$ are finite for $i$~$=$~$1$ and
$2$ despite individual graphs being divergent. Although this provides a check 
on the computation we have carried out a more stringent check in that the 
Green's function has to satisfy the Slavnov-Taylor identity underlying the 
charge current conservation. In practical terms this means that 
$\Sigma^{(1)}_{\bar{\psi} \gamma^\mu \psi}(p)$ must be equivalent to the quark 
$2$-point function after its renormalization in the same scheme. We have 
verified explicitly that this is indeed the case in the $\MSbar$ scheme. For 
the $\RI$ scheme agreement follows trivially as a result of the Slavnov-Taylor 
identity and the $\RI$ scheme condition defining $Z_\psi$ which is, \cite{1,2},
\begin{equation}
\left. \lim_{\epsilon \rightarrow 0} \left[
Z^{\mbox{\footnotesize{RI$^\prime$}}}_\psi \Sigma_\psi(p) \right]
\right|_{p^2 \, = \, \mu^2} ~=~ \pslash 
\end{equation}
and produces
\begin{equation}
\Sigma^{(1)\,\RIs}_{\bar{\psi} \gamma^\mu \psi}(p) ~=~ 1 ~+~ O(a^5)
\end{equation}
for all $\alpha$ and colour groups to the order we have calculated to. By 
contrast the $\MSbar$ expression is
\begin{eqnarray}
\left. \Sigma^{(1) ~ \MSbars}_{\bar{\psi}\gamma^\mu\psi}(p) \right|_{\alpha=0}
&=& 1
+ \left[
\frac{41}{4} C_A C_F
- \frac{7}{2} C_F T_F \Nf
- \frac{5}{8} C_F^2
- 3 \zeta_3 C_A C_F
\right] a^2
\nonumber \\
&&
+ \left[
\frac{1570}{81} C_F T_F^2 \Nf^2
+ \frac{159257}{648} C_A^2 C_F
- \frac{11887}{81} C_A C_F T_F \Nf
- \frac{3139}{24} \zeta_3 C_A^2 C_F
\right. \nonumber \\
&& \left. ~~~
- \frac{997}{24} C_A C_F^2
- \frac{79}{6} C_F^2 T_F \Nf
- \frac{73}{12} C_F^3
- \frac{69}{16} \zeta_4 C_A^2 C_F
+ \frac{52}{3} \zeta_3 C_A C_F T_F \Nf
\right. \nonumber \\
&& \left. ~~~
+ \frac{165}{4} \zeta_5 C_A^2 C_F
- 20 \zeta_5 C_A C_F^2
+ 6 \zeta_4 C_A C_F^2
+ 16 \zeta_3 C_F^2 T_F \Nf
\right. \nonumber \\
&& \left. ~~~
+ 44 \zeta_3 C_A C_F^2
\right] a^3
\nonumber \\
&&
+ \left[
400 \zeta_6 C_F^4
- \frac{5660951}{1296} C_A^2 C_F T_F \Nf
- \frac{1278281}{864} C_A^2 C_F^2
- \frac{1073621}{288} \zeta_3 C_A^3 C_F
\right. \nonumber \\
&& \left. ~~~
- \frac{447197}{432} C_A C_F^2 T_F \Nf
- \frac{264467}{384} \zeta_7 C_A^3 C_F
- \frac{63769}{72} \zeta_5 C_A^2 C_F T_F \Nf
\right. \nonumber \\
&& \left. ~~~
- \frac{42945}{128} C_F^4
- \frac{21391}{243} C_F T_F^3 \Nf^3
- \frac{20725}{256} \zeta_6 C_A^3 C_F
- \frac{18627}{512} \zeta_4 C_A^3 C_F
\right. \nonumber \\
&& \left. ~~~
- \frac{10185}{2} \zeta_7 C_A C_F^3
- \frac{4697}{8} \zeta_7 \frac{d_F^{abcd} d_A^{abcd}}{\NF}
- \frac{4275}{64} \zeta_6 \frac{d_F^{abcd} d_A^{abcd}}{\NF}
- \frac{3889}{2} \zeta_3 C_F^4
\right. \nonumber \\
&& \left. ~~~
- \frac{2263}{3} \zeta_3 C_A^2 C_F^2
- \frac{1263}{16} \zeta_4 C_A^2 C_F^2
- \frac{1225}{3} \zeta_5 C_A C_F^3
- \frac{796}{3} \zeta_3 C_F^2 T_F^2 \Nf^2
\right. \nonumber \\
&& \left. ~~~
- \frac{455}{2} \frac{d_F^{abcd} d_A^{abcd}}{\NF}
- \frac{441}{2} \zeta_7 C_A^2 C_F T_F \Nf
- \frac{351}{4} \zeta_3^2 C_A^2 C_F^2
\right. \nonumber \\
&& \left. ~~~
- \frac{280}{3} \zeta_3 C_A C_F T_F^2 \Nf^2
- \frac{16}{9} \zeta_3 C_F T_F^3 \Nf^3
- \frac{9}{16} \zeta_4 C_A^2 C_F T_F \Nf
\right. \nonumber \\
&& \left. ~~~
+ \frac{20}{3} \zeta_5 C_F^3 T_F \Nf
+ \frac{440}{9} \zeta_5 C_A C_F T_F^2 \Nf^2
+ \frac{613}{16} \zeta_3 \frac{d_F^{abcd} d_A^{abcd}}{\NF}
\right. \nonumber \\
&& \left. ~~~
+ \frac{779}{6} \zeta_3 C_A C_F^2 T_F \Nf
+ \frac{1669}{8} \zeta_5 C_A^2 C_F^2
+ \frac{3339}{64} \zeta_4 \frac{d_F^{abcd} d_A^{abcd}}{\NF}
\right. \nonumber \\
&& \left. ~~~
+ \frac{3925}{8} \zeta_6 C_A^2 C_F^2
+ \frac{4929}{32} \zeta_3^2 \frac{d_F^{abcd} d_A^{abcd}}{\NF}
+ \frac{10795}{64} \zeta_3^2 C_A^3 C_F
\right. \nonumber \\
&& \left. ~~~
+ \frac{12415}{32} \zeta_5 \frac{d_F^{abcd} d_A^{abcd}}{\NF}
+ \frac{21775}{144} C_F^3 T_F \Nf
+ \frac{27895}{16} \zeta_7 C_A^2 C_F^2
\right. \nonumber \\
&& \left. ~~~
+ \frac{32359}{8} \zeta_3 C_A C_F^3
+ \frac{46643}{24} \zeta_3 C_A^2 C_F T_F \Nf
+ \frac{79531}{216} C_F^2 T_F^2 \Nf^2
\right. \nonumber \\
&& \left. ~~~
+ \frac{124783}{108} C_A C_F T_F^2 \Nf^2
+ \frac{278975}{288} C_A C_F^3
+ \frac{2256031}{1152} \zeta_5 C_A^3 C_F
\right. \nonumber \\
&& \left. ~~~
+ \frac{164005363}{31104} C_A^3 C_F
- 1240 \zeta_5 \Nf \frac{d_F^{abcd} d_F^{abcd}}{\NF}
- 970 \zeta_5 C_F^4
- 900 \zeta_6 C_A C_F^3
\right. \nonumber \\
&& \left. ~~~
- 384 \zeta_3^2 \Nf \frac{d_F^{abcd} d_F^{abcd}}{\NF}
- 160 \zeta_3^2 C_A C_F^2 T_F \Nf
- 150 \zeta_4 C_F^4
- 103 \zeta_3 C_F^3 T_F \Nf
\right. \nonumber \\
&& \left. ~~~
- 80 \zeta_3^2 C_F^4
- 50 \zeta_6 C_A^2 C_F T_F \Nf
- 24 \zeta_4 C_F^3 T_F \Nf
- 12 \zeta_4 C_F^2 T_F^2 \Nf^2
\right. \nonumber \\
&& \left. ~~~
+ 12 \zeta_4 C_A C_F T_F^2 \Nf^2
+ 84 \zeta_3^2 C_A^2 C_F T_F \Nf
+ 100 \zeta_6 C_A C_F^2 T_F \Nf
\right. \nonumber \\
&& \left. ~~~
+ 118 \zeta_5 C_A C_F^2 T_F \Nf
+ 156 \zeta_3^2 C_A C_F^3
+ 318 \zeta_4 C_A C_F^3
+ 474 \Nf \frac{d_F^{abcd} d_F^{abcd}}{\NF}
\right. \nonumber \\
&& \left. ~~~
+ 882 \zeta_7 C_A C_F^2 T_F \Nf
+ 1124 \zeta_3 \Nf \frac{d_F^{abcd} d_F^{abcd}}{\NF}
+ 3528 \zeta_7 C_F^4
\right] a^4
\nonumber \\
&& +~ O(a^5)
\end{eqnarray}
in the Landau gauge. For completeness and to assist with the extraction of the 
relevant part of the Green's function for lattice matching we have
\begin{eqnarray}
\left. \Sigma^{(2) ~ \MSbars}_{\bar{\psi}\gamma^\mu\psi}(p) \right|_{\alpha=0}
&=& 
\left[
3 C_F^2
+ 4 C_F T_F \Nf
- \frac{25}{2} C_A C_F
\right] a^2
\nonumber \\
&&
+ \left[
\frac{242}{3} C_A C_F^2
+ \frac{245}{4} \zeta_3 C_A^2 C_F
+ \frac{1528}{9} C_A C_F T_F \Nf
- \frac{19979}{72} C_A^2 C_F
\right. \nonumber \\
&& \left. ~~~
- \frac{208}{9} C_F T_F^2 \Nf^2
- \frac{28}{3} C_F^2 T_F \Nf
- 24 \zeta_3 C_A C_F^2
- 16 \zeta_3 C_A C_F T_F \Nf
\right. \nonumber \\
&& \left. ~~~
- 3 C_F^3
\right] a^3
\nonumber \\
&&
+ \left[
\frac{2039}{8} C_F^4
- \frac{761141}{108} C_A^3 C_F
- \frac{32773}{48} C_A C_F^3
- \frac{15830}{9} C_A C_F T_F^2 \Nf^2
\right. \nonumber \\
&& \left. ~~~
- \frac{10375}{16} \zeta_5 C_A^3 C_F
- \frac{4514}{3} \zeta_3 C_A^2 C_F T_F \Nf
- \frac{2186}{9} C_F^2 T_F^2 \Nf^2
\right. \nonumber \\
&& \left. ~~~
- \frac{1113}{4} \zeta_3 \frac{d_F^{abcd} d_A^{abcd}}{\NF}
- \frac{1019}{2} \zeta_3 C_A^2 C_F^2
- \frac{367}{3} C_F^3 T_F \Nf
\right. \nonumber \\
&& \left. ~~~
+ \frac{224}{3} \zeta_3 C_A C_F T_F^2 \Nf^2
+ \frac{855}{4} \zeta_5 \frac{d_F^{abcd} d_A^{abcd}}{\NF}
+ \frac{4000}{27} C_F T_F^3 \Nf^3
\right. \nonumber \\
&& \left. ~~~
+ \frac{7031}{36} C_A C_F^2 T_F \Nf
+ \frac{56764}{9} C_A^2 C_F T_F \Nf
+ \frac{154157}{72} C_A^2 C_F^2
\right. \nonumber \\
&& \left. ~~~
+ \frac{307915}{96} \zeta_3 C_A^3 C_F
- 1705 \zeta_3 C_A C_F^3
- 1280 \zeta_5 C_F^4
- 1130 \zeta_5 C_A^2 C_F^2
\right. \nonumber \\
&& \left. ~~~
- 480 \zeta_5 C_A C_F^2 T_F \Nf
- 256 \Nf \frac{d_F^{abcd} d_F^{abcd}}{\NF}
- 60 \zeta_3 C_A C_F^2 T_F \Nf
\right. \nonumber \\
&& \left. ~~~
+ 128 \zeta_3 C_F^3 T_F \Nf
+ 134 \frac{d_F^{abcd} d_A^{abcd}}{\NF}
+ 192 \zeta_3 C_F^2 T_F^2 \Nf^2
\right. \nonumber \\
&& \left. ~~~
+ 490 \zeta_5 C_A^2 C_F T_F \Nf
+ 800 \zeta_3 C_F^4
+ 2880 \zeta_5 C_A C_F^3
\right] a^4 ~+~ O(a^5) ~.
\end{eqnarray}
We also record
\begin{eqnarray}
\left. \Sigma^{(2) ~ \RIs}_{\bar{\psi}\gamma^\mu\psi}(p) \right|_{\alpha=0} &=& 
\left[
3 C_F^2
+ 4 C_F T_F \Nf
- \frac{25}{2} C_A C_F
\right] a^2
\nonumber \\
&&
+ \left[
\frac{242}{3} C_A C_F^2
+ \frac{245}{4} \zeta_3 C_A^2 C_F
+ \frac{1528}{9} C_A C_F T_F \Nf
- \frac{19979}{72} C_A^2 C_F
\right. \nonumber \\
&& \left. ~~~
- \frac{208}{9} C_F T_F^2 \Nf^2
- \frac{28}{3} C_F^2 T_F \Nf
- 24 \zeta_3 C_A C_F^2
- 16 \zeta_3 C_A C_F T_F \Nf
\right. \nonumber \\
&& \left. ~~~
- 3 C_F^3
\right] a^3
\nonumber \\
&&
+ \left[
\frac{1027}{4} C_F^4
- \frac{761141}{108} C_A^3 C_F
- \frac{15830}{9} C_A C_F T_F^2 \Nf^2
- \frac{10375}{16} \zeta_5 C_A^3 C_F
\right. \nonumber \\
&& \left. ~~~
- \frac{4514}{3} \zeta_3 C_A^2 C_F T_F \Nf
- \frac{2164}{3} C_A C_F^3
- \frac{2060}{9} C_F^2 T_F^2 \Nf^2
\right. \nonumber \\
&& \left. ~~~
- \frac{1113}{4} \zeta_3 \frac{d_F^{abcd} d_A^{abcd}}{\NF}
- \frac{328}{3} C_F^3 T_F \Nf
+ \frac{224}{3} \zeta_3 C_A C_F T_F^2 \Nf^2
\right. \nonumber \\
&& \left. ~~~
+ \frac{855}{4} \zeta_5 \frac{d_F^{abcd} d_A^{abcd}}{\NF}
+ \frac{995}{9} C_A C_F^2 T_F \Nf
+ \frac{4000}{27} C_F T_F^3 \Nf^3
\right. \nonumber \\
&& \left. ~~~
+ \frac{56764}{9} C_A^2 C_F T_F \Nf
+ \frac{81691}{36} C_A^2 C_F^2
+ \frac{307915}{96} \zeta_3 C_A^3 C_F
\right. \nonumber \\
&& \left. ~~~
- 1696 \zeta_3 C_A C_F^3
- 1280 \zeta_5 C_F^4
- 1130 \zeta_5 C_A^2 C_F^2
- 547 \zeta_3 C_A^2 C_F^2
\right. \nonumber \\
&& \left. ~~~
- 480 \zeta_5 C_A C_F^2 T_F \Nf
- 256 \Nf \frac{d_F^{abcd} d_F^{abcd}}{\NF}
- 48 \zeta_3 C_A C_F^2 T_F \Nf
\right. \nonumber \\
&& \left. ~~~
+ 128 \zeta_3 C_F^3 T_F \Nf
+ 134 \frac{d_F^{abcd} d_A^{abcd}}{\NF}
+ 192 \zeta_3 C_F^2 T_F^2 \Nf^2
\right. \nonumber \\
&& \left. ~~~
+ 490 \zeta_5 C_A^2 C_F T_F \Nf
+ 800 \zeta_3 C_F^4
+ 2880 \zeta_5 C_A C_F^3
\right] a^4 ~+~ O(a^5)
\end{eqnarray}
to complete the analysis in both schemes. We note that $\left. 
\Sigma^{(2) ~ \RIs}_{\bar{\psi}\gamma^\mu\psi}(p) \right|_{\alpha=0}$ differs 
from its $\MSbar$ counterpart at four loops. The difference between the four
loop coefficients in each expression is related to the product of the 
coefficient of the two loop Landau gauge terms of $Z_\psi^{\RIs}$ and $\left. 
\Sigma^{(1) ~ \MSbars}_{\bar{\psi}\gamma^\mu\psi}(p) \right|_{\alpha=0}$. Full 
expressions for $\Sigma^{(i)}_{\bar{\psi}\gamma^\mu\psi}(p)$ in both schemes 
for an arbitrary linear covariant gauge are included in the associated data 
file. Finally we record the numerical values are
\begin{eqnarray}
\left. \Sigma^{(1) ~ \MSbars}_{\bar{\psi}\gamma^\mu\psi}(p) 
\right|_{\alpha=0}^{SU(3)} &=& 
1 + [25.464206 - 2.333333 \Nf] a^2 
\nonumber \\
&&
+~ [6.460905 \Nf^2 - 246.442650 \Nf + 1489.980500] a^3 
\nonumber \\
&&
+~ [78097.273944 - 18600.055317 \Nf + 1123.006264 \Nf^2 
- 15.027633 \Nf^3] a^4 
\nonumber \\
&& +~ O(a^5)
\nonumber \\
\left. \Sigma^{(2) ~ \MSbars}_{\bar{\psi}\gamma^\mu\psi}(p) 
\right|_{\alpha=0}^{SU(3)} &=& 
[2.666667 \Nf - 44.666667] a^2 
\nonumber \\
&&
+~ [ 292.793438 \Nf - 7.703704 \Nf^2 - 2177.073682] a^3 
\nonumber \\
&&
+~ [24.691358 \Nf^3 - 1674.510402 \Nf^2 + 29042.096003 \Nf
- 131900.241538] a^4 
\nonumber \\
&& +~ O(a^5)
\nonumber \\
\left. \Sigma^{(2) ~ \RIs}_{\bar{\psi}\gamma^\mu\psi}(p) 
\right|_{\alpha=0}^{SU(3)} &=& 
[2.666667 \Nf - 44.666667] a^2 
\nonumber \\
&&
+~ [292.793438 \Nf - 7.703704 \Nf^2 - 2177.073682] a^3 
\nonumber \\
&&
+~ [24.691358 \Nf^3 - 1668.288179 \Nf^2 + 28869.969231 \Nf 
- 130762.840335] a^4 
\nonumber \\
&& +~ O(a^5)
\end{eqnarray}
when the colour group is $SU(3)$. To gauge the effect the four loop term has we
have again evaluated the two $\MSbar$ scheme expressions at 
$\alpha_s$~$=$~$0.12$ when $\Nf$~$=$~$3$. For 
$\left. \Sigma^{(1) ~ \MSbars}_{\bar{\psi}\gamma^\mu\psi}(p) 
\right|_{\alpha=0}^{SU(3)}$ we note that the value at the three successive loop
orders are $1.001684$, $1.002388$ and $1.002654$ as there is no one loop term
while the respective values are $-$~$0.003344$, $-$~$0.004535$ and
$-$~$0.005027$ for $\left. \Sigma^{(2) ~ \MSbars}_{\bar{\psi}\gamma^\mu\psi}(p)
\right|_{\alpha=0}^{SU(3)}$. For 
$\left. \Sigma^{(2) ~ \RIs}_{\bar{\psi}\gamma^\mu\psi}(p)
\right|_{\alpha=0}^{SU(3)}$ the two and three loop values are the same as the
$\MSbar$ ones while the four loop value is $-$~$0.005021$. Clearly there is a 
small discrepancy between the three and four loop $\MSbar$ values for channel 
$1$ which suggests that there is a degree of convergence. 
 
\sect{Discussion.}

We have now extended the renormalization group functions of QCD in the 
$\RI$ scheme to five loops. In addition the $2$-point functions of the three
fields are also available to four loops in the same scheme. However the results
of more immediate use provided in this article are the determination of the 
four loop Green's functions where the quark mass operator and separately the 
vector current are inserted at zero momentum in a quark $2$-point function in 
both the $\MSbar$ and $\RI$ schemes. These are important for the wider and 
ongoing lattice gauge theory programme of measuring quark masses more 
accurately. Our observation is that the four loop corrections of these operator
Green's functions are not significantly different from their three loop values 
at a particular reference point. This should in principle allow for a better
understanding of errors in extrapolating and matching lattice data to high 
energy for the exceptional momentum configuration considered here. We recall
that the corresponding non-exceptional point symmetric point renormalization 
was carried out to three loops in \cite{47} in the Landau gauge. While this
equated to the loop order achieved in \cite{13}, various analyses such as that
of \cite{48} used those results to evaluate operator renormalization constants
on the lattice. For instance, \cite{48} confirmed the expected behaviour of the
renormalization constants over a wide range of momenta down to infrared scales.
At a particular point it turned out that the exceptional momentum case began to
deviate from expectations. As the three loop $\RI$ perturbative renormalization
was used for that it would be interesting to see whether the four loop 
information of this study improves the behaviour in the infrared and if so then
how well does it compare with the symmetric point measurements based on both 
the two, \cite{13}, and three loop Landau gauge data, \cite{47}.

\vspace{1cm}
\noindent
{\bf Acknowledgements.} The author thanks R.H. Mason for useful discussions.
This work was carried out with the support of the STFC through the Consolidated
Grant ST/T000988/1. For the purpose of open access, the author has applied a 
Creative Commons Attribution (CC-BY) licence to any Author Accepted Manuscript 
version arising. The data representing the main results here are accessible in 
electronic form from the arXiv ancillary directory associated with this 
article.

\appendix

\sect{Expressions for a general Lie group.}

As the renormalization group functions for a general colour group are of 
interest, for example with respect to the Casimir location and dependence on 
the number of colours, we record the various anomalous dimensions in the
Landau gauge for the $\RI$ scheme. First the expressions for the field
anomalous dimensions are 
\begin{eqnarray}
\gamma_A^{\RIs}(a,0) &=&
\left[
\frac{4}{3} T_F \Nf
- \frac{13}{6} C_A
\right] a
+ \left[
4 C_F T_F \Nf
- \frac{3727}{216} C_A^2
- \frac{80}{27} T_F^2 \Nf^2
+ \frac{452}{27} C_A T_F \Nf
\right] a^2 
\nonumber \\
&&
+ \left[
\frac{26054}{81} C_A^2 T_F \Nf
- \frac{2127823}{7776} C_A^3
- \frac{290}{3} C_A T_F^2 \Nf^2
- \frac{280}{3} \zeta_3 C_A C_F T_F \Nf
\right. \nonumber \\
&& \left. ~~~~
- \frac{188}{3} C_F T_F^2 \Nf^2
- \frac{64}{3} \zeta_3 C_A T_F^2 \Nf^2
+ \frac{98}{3} \zeta_3 C_A^2 T_F \Nf
+ \frac{128}{3} \zeta_3 C_F T_F^2 \Nf^2
\right. \nonumber \\
&& \left. ~~~~
+ \frac{291}{2} C_A C_F T_F \Nf
+ \frac{361}{16} \zeta_3 C_A^3
+ \frac{1600}{243} T_F^3 \Nf^3
- 2 C_F^2 T_F \Nf
\right] a^3 
\nonumber \\
&&
+ \left[
192 \zeta_3 C_A T_F^3 \Nf^3
- \frac{3012081353}{559872} C_A^4
- \frac{4608833}{1458} C_A^2 T_F^2 \Nf^2
\right. \nonumber \\
&& \left. ~~~~
- \frac{821905}{243} C_A C_F T_F^2 \Nf^2
- \frac{32000}{2187} T_F^4 \Nf^4
- \frac{29905}{48} \zeta_5 C_A^3 T_F \Nf
\right. \nonumber \\
&& \left. ~~~~
- \frac{21920}{9} \zeta_3 C_A^2 C_F T_F \Nf
- \frac{10185}{64} \zeta_5 \frac{d_A^{abcd} d_A^{abcd}}{\NA}
- \frac{8788}{9} \zeta_3 C_A C_F^2 T_F \Nf
\right. \nonumber \\
&& \left. ~~~~
- \frac{7498}{9} \zeta_3 C_A^2 T_F^2 \Nf^2
- \frac{4160}{27} C_A C_F^2 T_F \Nf
- \frac{3712}{9} \zeta_3 C_F T_F^3 \Nf^3
\right. \nonumber \\
&& \left. ~~~~
- \frac{1760}{27} C_F^2 T_F^2 \Nf^2
- \frac{989}{12} \zeta_3 \frac{d_A^{abcd} d_A^{abcd}}{\NA}
- \frac{512}{3} \zeta_3 \Nf^2 \frac{d_F^{abcd} d_F^{abcd}}{\NA}
\right. \nonumber \\
&& \left. ~~~~
- \frac{512}{9} \Nf \frac{d_F^{abcd} d_A^{abcd}}{\NA}
+ \frac{640}{3} \zeta_5 C_A^2 T_F^2 \Nf^2
+ \frac{659}{144} \frac{d_A^{abcd} d_A^{abcd}}{\NA}
+ \frac{704}{9} \Nf^2 \frac{d_F^{abcd} d_F^{abcd}}{\NA}
\right. \nonumber \\
&& \left. ~~~~
+ \frac{1376}{3} \zeta_3 \Nf \frac{d_F^{abcd} d_A^{abcd}}{\NA}
+ \frac{5408}{9} \zeta_3 C_F^2 T_F^2 \Nf^2
+ \frac{5968}{3} \zeta_3 C_A C_F T_F^2 \Nf^2
\right. \nonumber \\
&& \left. ~~~~
+ \frac{8999}{36} \zeta_3 C_A^3 T_F \Nf
+ \frac{134632}{243} C_F T_F^3 \Nf^3
+ \frac{475205}{1536} \zeta_5 C_A^4
+ \frac{716429}{768} \zeta_3 C_A^4
\right. \nonumber \\
&& \left. ~~~~
+ \frac{956764}{2187} C_A T_F^3 \Nf^3
+ \frac{2130347}{486} C_A^2 C_F T_F \Nf
+ \frac{529941473}{69984} C_A^3 T_F \Nf
\right. \nonumber \\
&& \left. ~~~~
- 760 \zeta_5 C_A^2 C_F T_F \Nf
- 640 \zeta_5 C_F^2 T_F^2 \Nf^2
- 46 C_F^3 T_F \Nf
\right. \nonumber \\
&& \left. ~~~~
+ 120 \zeta_5 \Nf \frac{d_F^{abcd} d_A^{abcd}}{\NA}
+ 320 \zeta_5 C_A C_F T_F^2 \Nf^2
+ 1520 \zeta_5 C_A C_F^2 T_F \Nf
\right] a^4 
\nonumber \\
&&
+ \left[
\frac{28974184619}{995328} \zeta_3 C_A^5
+ \frac{4892001875}{110592} \zeta_5 C_A^5
- \frac{5298101107831}{40310784} C_A^5
\right. \nonumber \\
&& \left. ~~~~
- \frac{18075016135}{157464} C_A^3 T_F^2 \Nf^2
- \frac{5939172295}{294912} \zeta_7 C_A^5
+ \frac{1091693262941}{5038848} C_A^4 T_F \Nf
\right. \nonumber \\
&& \left. ~~~~
- \frac{790040875}{10368} \zeta_5 C_A^4 T_F \Nf
- \frac{674419927}{62208} \zeta_3 C_A^4 T_F \Nf
\right. \nonumber \\
&& \left. ~~~~
- \frac{221064985}{1458} C_A^2 C_F T_F^2 \Nf^2
- \frac{199659583}{110592} \zeta_3^2 C_A^5
\right. \nonumber \\
&& \left. ~~~~
- \frac{140495179}{13824} \zeta_3 C_A \frac{d_A^{abcd} d_A^{abcd}}{\NA}
- \frac{121750363}{12288} \zeta_7 C_A \frac{d_A^{abcd} d_A^{abcd}}{\NA}
\right. \nonumber \\
&& \left. ~~~~
- \frac{68513455}{1296} \zeta_3 C_A^3 C_F T_F \Nf
- \frac{32911972}{19683} C_A T_F^4 \Nf^4
- \frac{15005047}{648} \zeta_3 C_A^3 T_F^2 \Nf^2
\right. \nonumber \\
&& \left. ~~~~
- \frac{4571425}{768} \zeta_5 C_A \frac{d_A^{abcd} d_A^{abcd}}{\NA}
- \frac{3031633}{54} \zeta_3 C_A^2 C_F^2 T_F \Nf
- \frac{963825}{16} \zeta_5 C_A^3 C_F T_F \Nf
\right. \nonumber \\
&& \left. ~~~~
- \frac{951901}{324} C_A C_F^2 T_F^2 \Nf^2
- \frac{677390}{81} C_A^2 C_F^2 T_F \Nf
- \frac{591424}{27} \zeta_3 C_A C_F T_F^3 \Nf^3
\right. \nonumber \\
&& \left. ~~~~
- \frac{453040}{9} \zeta_5 C_A C_F^2 T_F^2 \Nf^2
- \frac{442757}{36} \zeta_7 C_A^3 C_F T_F \Nf
\right. \nonumber \\
&& \left. ~~~~
- \frac{429088}{27} \zeta_3 T_F \Nf^2 \frac{d_F^{abcd} d_A^{abcd}}{\NA}
- \frac{385745}{72} \zeta_3^2 T_F \Nf \frac{d_A^{abcd} d_A^{abcd}}{\NA}
\right. \nonumber \\
&& \left. ~~~~
- \frac{368000}{27} \zeta_5 C_A C_F T_F^3 \Nf^3
- \frac{338560}{243} \zeta_3 C_A T_F^4 \Nf^4
\right. \nonumber \\
&& \left. ~~~~
- \frac{258632}{27} \zeta_3 C_A \Nf^2 \frac{d_F^{abcd} d_F^{abcd}}{\NA}
- \frac{227363}{9} \zeta_3^2 C_A \Nf \frac{d_F^{abcd} d_A^{abcd}}{\NA}
\right. \nonumber \\
&& \left. ~~~~
- \frac{121352}{27} C_F T_F^4 \Nf^4
- \frac{100040}{27} C_A \Nf \frac{d_F^{abcd} d_A^{abcd}}{\NA}
- \frac{88480}{3} \zeta_7 C_A C_F^3 T_F \Nf
\right. \nonumber \\
&& \left. ~~~~
- \frac{58339}{576} \zeta_3^2 C_A^4 T_F \Nf
- \frac{52960}{9} \zeta_5 T_F \Nf^2 \frac{d_F^{abcd} d_A^{abcd}}{\NA}
- \frac{40480}{9} \zeta_5 C_A^2 T_F^3 \Nf^3
\right. \nonumber \\
&& \left. ~~~~
- \frac{35840}{3} \zeta_7 C_A C_F^2 T_F^2 \Nf^2
- \frac{20224}{9} \zeta_3^2 C_A \Nf^2 \frac{d_F^{abcd} d_F^{abcd}}{\NA}
\right. \nonumber \\
&& \left. ~~~~
- \frac{17344}{9} T_F \Nf^3 \frac{d_F^{abcd} d_F^{abcd}}{\NA}
- \frac{8069}{2} C_A C_F^3 T_F \Nf
- \frac{7928}{9} \zeta_3^2 C_A^3 T_F^2 \Nf^2
\right. \nonumber \\
&& \left. ~~~~
- \frac{7877}{72} T_F \Nf \frac{d_A^{abcd} d_A^{abcd}}{\NA}
- \frac{6400}{3} \zeta_5 T_F \Nf^3 \frac{d_F^{abcd} d_F^{abcd}}{\NA}
- \frac{5120}{3} \zeta_3 C_F \Nf^2 \frac{d_F^{abcd} d_F^{abcd}}{\NA}
\right. \nonumber \\
&& \left. ~~~~
- \frac{4160}{3} C_F \Nf^2 \frac{d_F^{abcd} d_F^{abcd}}{\NA}
- \frac{3580}{3} \zeta_5 C_F \Nf \frac{d_F^{abcd} d_A^{abcd}}{\NA}
- \frac{1588}{3} \zeta_3 C_A C_F^3 T_F \Nf
\right. \nonumber \\
&& \left. ~~~~
- \frac{1544}{3} \zeta_3 C_F \Nf \frac{d_F^{abcd} d_A^{abcd}}{\NA}
- \frac{1274}{3} \zeta_7 C_A \Nf \frac{d_F^{abcd} d_A^{abcd}}{\NA}
- \frac{832}{3} \zeta_4 T_F \Nf^2 \frac{d_F^{abcd} d_A^{abcd}}{\NA}
\right. \nonumber \\
&& \left. ~~~~
- \frac{707}{18} \zeta_4 C_A^4 T_F \Nf
- \frac{692}{9} \zeta_4 C_A^2 C_F T_F^2 \Nf^2
- \frac{572}{3} \zeta_4 C_A \frac{d_A^{abcd} d_A^{abcd}}{\NA}
\right. \nonumber \\
&& \left. ~~~~
- \frac{416}{3} \zeta_4 C_A \Nf^2 \frac{d_F^{abcd} d_F^{abcd}}{\NA}
- \frac{392}{3} \zeta_7 T_F \Nf^2 \frac{d_F^{abcd} d_A^{abcd}}{\NA}
- \frac{352}{9} \zeta_4 C_F^2 T_F^3 \Nf^3
\right. \nonumber \\
&& \left. ~~~~
- \frac{286}{9} \zeta_4 C_A^2 C_F^2 T_F \Nf
+ \frac{22}{9} \zeta_4 C_A^3 T_F^2 \Nf^2
+ \frac{112}{9} \zeta_4 C_A^2 T_F^3 \Nf^3
+ \frac{143}{36} \zeta_4 C_A^5
\right. \nonumber \\
&& \left. ~~~~
+ \frac{224}{9} \zeta_4 C_A C_F T_F^3 \Nf^3
+ \frac{256}{3} \zeta_4 T_F \Nf^3 \frac{d_F^{abcd} d_F^{abcd}}{\NA}
+ \frac{352}{3} \zeta_4 T_F \Nf \frac{d_A^{abcd} d_A^{abcd}}{\NA}
\right. \nonumber \\
&& \left. ~~~~
+ \frac{533}{9} \zeta_4 C_A^3 C_F T_F \Nf
+ \frac{748}{9} \zeta_4 C_A C_F^2 T_F^2 \Nf^2
+ \frac{944}{3} C_F \Nf \frac{d_F^{abcd} d_A^{abcd}}{\NA}
\right. \nonumber \\
&& \left. ~~~~
+ \frac{1352}{3} \zeta_4 C_A \Nf \frac{d_F^{abcd} d_A^{abcd}}{\NA}
+ \frac{4157}{6} C_F^4 T_F \Nf
+ \frac{6784}{3} \zeta_3^2 C_A^2 C_F T_F^2 \Nf^2
\right. \nonumber \\
&& \left. ~~~~
+ \frac{8192}{9} \zeta_3^2 T_F \Nf^3 \frac{d_F^{abcd} d_F^{abcd}}{\NA}
+ \frac{10240}{81} \zeta_5 C_A T_F^4 \Nf^4
+ \frac{10526}{9} C_F^3 T_F^2 \Nf^2
\right. \nonumber \\
&& \left. ~~~~
+ \frac{12800}{3} \zeta_5 C_F \Nf^2 \frac{d_F^{abcd} d_F^{abcd}}{\NA}
+ \frac{12896}{9} T_F \Nf^2 \frac{d_F^{abcd} d_A^{abcd}}{\NA}
+ \frac{15872}{27} \zeta_3^2 C_A^2 T_F^3 \Nf^3
\right. \nonumber \\
&& \left. ~~~~
+ \frac{17665}{3} \zeta_3^2 C_A^3 C_F T_F \Nf
+ \frac{20480}{27} \zeta_5 C_F T_F^4 \Nf^4
+ \frac{35680}{9} \zeta_5 C_A \Nf^2 \frac{d_F^{abcd} d_F^{abcd}}{\NA}
\right. \nonumber \\
&& \left. ~~~~
+ \frac{35840}{3} \zeta_7 C_F^3 T_F^2 \Nf^2
+ \frac{44800}{9} \zeta_7 C_A^2 C_F T_F^2 \Nf^2
+ \frac{82240}{27} \zeta_3 T_F \Nf^3 \frac{d_F^{abcd} d_F^{abcd}}{\NA}
\right. \nonumber \\
&& \left. ~~~~
+ \frac{86464}{9} \zeta_3^2 T_F \Nf^2 \frac{d_F^{abcd} d_A^{abcd}}{\NA}
+ \frac{88480}{3} \zeta_7 C_A^2 C_F^2 T_F \Nf
\right. \nonumber \\
&& \left. ~~~~
+ \frac{99775}{9} \zeta_5 C_A \Nf \frac{d_F^{abcd} d_A^{abcd}}{\NA}
+ \frac{102400}{9} \zeta_5 C_F^2 T_F^3 \Nf^3
+ \frac{131467}{24} \zeta_7 T_F \Nf \frac{d_A^{abcd} d_A^{abcd}}{\NA}
\right. \nonumber \\
&& \left. ~~~~
+ \frac{163784}{27} C_A \Nf^2 \frac{d_F^{abcd} d_F^{abcd}}{\NA}
+ \frac{223186}{81} C_F^2 T_F^3 \Nf^3
+ \frac{224195}{144} \zeta_5 T_F \Nf \frac{d_A^{abcd} d_A^{abcd}}{\NA}
\right. \nonumber \\
&& \left. ~~~~
+ \frac{237440}{81} \zeta_3 C_F T_F^4 \Nf^4
+ \frac{483430}{9} \zeta_5 C_A^2 C_F^2 T_F \Nf
+ \frac{522002}{9} \zeta_3 C_A^2 C_F T_F^2 \Nf^2
\right. \nonumber \\
&& \left. ~~~~
+ \frac{640000}{19683} T_F^5 \Nf^5
+ \frac{745015}{3456} C_A \frac{d_A^{abcd} d_A^{abcd}}{\NA}
+ \frac{1098749}{27} \zeta_3 C_A \Nf \frac{d_F^{abcd} d_A^{abcd}}{\NA}
\right. \nonumber \\
&& \left. ~~~~
+ \frac{1488560}{27} \zeta_5 C_A^2 C_F T_F^2 \Nf^2
+ \frac{1615204}{27} \zeta_3 C_A C_F^2 T_F^2 \Nf^2
+ \frac{2995112}{243} \zeta_3 C_A^2 T_F^3 \Nf^3
\right. \nonumber \\
&& \left. ~~~~
+ \frac{3683755}{108} \zeta_5 C_A^3 T_F^2 \Nf^2
+ \frac{3835049}{864} \zeta_3 T_F \Nf \frac{d_A^{abcd} d_A^{abcd}}{\NA}
+ \frac{35251481}{729} C_A C_F T_F^3 \Nf^3
\right. \nonumber \\
&& \left. ~~~~
+ \frac{46576943}{1152} \zeta_7 C_A^4 T_F \Nf
+ \frac{66141673}{4608} \zeta_3^2 C_A \frac{d_A^{abcd} d_A^{abcd}}{\NA}
+ \frac{925396565}{39366} C_A^2 T_F^3 \Nf^3
\right. \nonumber \\
&& \left. ~~~~
+ \frac{1619234591}{11664} C_A^3 C_F T_F \Nf
- 14400 \zeta_3 C_F^2 T_F^3 \Nf^3
- 13920 \zeta_5 C_F^3 T_F^2 \Nf^2
\right. \nonumber \\
&& \left. ~~~~
- 12628 \zeta_7 C_A^3 T_F^2 \Nf^2
- 4864 \zeta_3^2 C_A C_F^2 T_F^2 \Nf^2
- 2048 \zeta_3^2 C_A C_F T_F^3 \Nf^3
\right. \nonumber \\
&& \left. ~~~~
- 476 \zeta_7 C_F \Nf \frac{d_F^{abcd} d_A^{abcd}}{\NA}
- 288 \zeta_3^2 C_F \Nf \frac{d_F^{abcd} d_A^{abcd}}{\NA}
+ 128 \zeta_3 C_F^4 T_F \Nf
\right. \nonumber \\
&& \left. ~~~~
+ 1280 \zeta_3 C_F^3 T_F^2 \Nf^2
+ 2048 \zeta_3^2 C_F^2 T_F^3 \Nf^3
+ 28300 \zeta_5 C_A C_F^3 T_F \Nf
\right] a^5
\nonumber \\
&& +~ O(a^6)
\end{eqnarray}
\begin{eqnarray}
\gamma_c^{\RIs}(a,0) &=&
-~ \frac{3}{4} C_A a
+ \left[
\frac{13}{6} C_A T_F \Nf
- \frac{271}{48} C_A^2
\right] a^2 
\nonumber \\
&&
+ \left[
\frac{13}{2} \zeta_3 C_A^2 T_F \Nf
- \frac{157303}{1728} C_A^3
- \frac{250}{27} C_A T_F^2 \Nf^2
+ \frac{61}{4} C_A C_F T_F \Nf
+ \frac{211}{32} \zeta_3 C_A^3
\right. \nonumber \\
&& \left. ~~~~
+ \frac{13445}{216} C_A^2 T_F \Nf
- 12 \zeta_3 C_A C_F T_F \Nf
\right] a^3 
\nonumber \\
&&
+ \left[
\frac{2825}{48} \zeta_3 C_A^3 T_F \Nf
- \frac{219440027}{124416} C_A^4
- \frac{82325}{162} C_A^2 T_F^2 \Nf^2
- \frac{10295}{54} C_A C_F T_F^2 \Nf^2
\right. \nonumber \\
&& \left. ~~~~
- \frac{675}{2} \zeta_3 C_A^2 C_F T_F \Nf
- \frac{609}{8} \zeta_3 \frac{d_A^{abcd} d_A^{abcd}}{\NA}
- \frac{505}{8} \zeta_5 C_A^3 T_F \Nf
- \frac{295}{12} C_A C_F^2 T_F \Nf
\right. \nonumber \\
&& \left. ~~~~
- \frac{158}{3} \zeta_3 C_A^2 T_F^2 \Nf^2
+ \frac{69}{32} \frac{d_A^{abcd} d_A^{abcd}}{\NA}
+ \frac{10185}{128} \zeta_5 \frac{d_A^{abcd} d_A^{abcd}}{\NA}
+ \frac{10820}{243} C_A T_F^3 \Nf^3
\right. \nonumber \\
&& \left. ~~~~
+ \frac{70545}{1024} \zeta_5 C_A^4
+ \frac{123715}{216} C_A^2 C_F T_F \Nf
+ \frac{365387}{1536} \zeta_3 C_A^4
+ \frac{9013883}{5184} C_A^3 T_F \Nf
\right. \nonumber \\
&& \left. ~~~~
- 74 \zeta_3 C_A C_F^2 T_F \Nf
- 60 \zeta_5 \Nf \frac{d_F^{abcd} d_A^{abcd}}{\NA}
- 60 \zeta_5 C_A^2 C_F T_F \Nf
\right. \nonumber \\
&& \left. ~~~~
+ 48 \zeta_3 \Nf \frac{d_F^{abcd} d_A^{abcd}}{\NA}
+ 120 \zeta_5 C_A C_F^2 T_F \Nf
+ 128 \zeta_3 C_A C_F T_F^2 \Nf^2
\right] a^4 
\nonumber \\
&&
+ \left[
\frac{60381388927}{1119744} C_A^4 T_F \Nf
- \frac{374313393145}{8957952} C_A^5
- \frac{695857939}{196608} \zeta_7 C_A^5
\right. \nonumber \\
&& \left. ~~~~
- \frac{68955821}{2916} C_A^3 T_F^2 \Nf^2
- \frac{17068475}{2304} \zeta_5 C_A^4 T_F \Nf
- \frac{8364239}{648} C_A^2 C_F T_F^2 \Nf^2
\right. \nonumber \\
&& \left. ~~~~
- \frac{7523471}{1024} \zeta_3 C_A \frac{d_A^{abcd} d_A^{abcd}}{\NA}
- \frac{3068531}{8192} \zeta_3^2 C_A^5
- \frac{2164553}{1536} \zeta_3 C_A^4 T_F \Nf
\right. \nonumber \\
&& \left. ~~~~
- \frac{596896}{2187} C_A T_F^4 \Nf^4
- \frac{448897}{1024} \zeta_3^2 C_A \frac{d_A^{abcd} d_A^{abcd}}{\NA}
- \frac{328191}{32} \zeta_3 C_A^3 C_F T_F \Nf
\right. \nonumber \\
&& \left. ~~~~
- \frac{317605}{144} \zeta_3 C_A^3 T_F^2 \Nf^2
- \frac{172253}{144} C_A^2 C_F^2 T_F \Nf
- \frac{164135}{96} \zeta_5 T_F \Nf \frac{d_A^{abcd} d_A^{abcd}}{\NA}
\right. \nonumber \\
&& \left. ~~~~
- \frac{77889}{64} \zeta_7 T_F \Nf \frac{d_A^{abcd} d_A^{abcd}}{\NA}
- \frac{42109}{384} \zeta_3^2 C_A^4 T_F \Nf
- \frac{40805}{6} \zeta_5 C_A \Nf \frac{d_F^{abcd} d_A^{abcd}}{\NA}
\right. \nonumber \\
&& \left. ~~~~
- \frac{33635}{8} \zeta_5 C_A^3 C_F T_F \Nf
- \frac{16681}{24} \zeta_7 C_A^3 C_F T_F \Nf
- \frac{12808}{3} \zeta_3 C_A^2 C_F^2 T_F \Nf
\right. \nonumber \\
&& \left. ~~~~
- \frac{6560}{3} \zeta_5 C_A C_F^2 T_F^2 \Nf^2
- \frac{5887}{24} C_A C_F^2 T_F^2 \Nf^2
- \frac{5504}{3} \zeta_3 T_F \Nf^2 \frac{d_F^{abcd} d_A^{abcd}}{\NA}
\right. \nonumber \\
&& \left. ~~~~
- \frac{3641}{48} T_F \Nf \frac{d_A^{abcd} d_A^{abcd}}{\NA}
- \frac{1333}{2} \zeta_3^2 C_A \Nf \frac{d_F^{abcd} d_A^{abcd}}{\NA}
- \frac{1292}{9} C_A \Nf \frac{d_F^{abcd} d_A^{abcd}}{\NA}
\right. \nonumber \\
&& \left. ~~~~
- \frac{680}{9} \zeta_5 C_A^2 T_F^3 \Nf^3
- \frac{422}{3} \zeta_3^2 C_A^3 T_F^2 \Nf^2
- \frac{320}{3} \zeta_5 C_A C_F T_F^3 \Nf^3
- \frac{51}{4} \zeta_4 C_A^4 T_F \Nf
\right. \nonumber \\
&& \left. ~~~~
+ \frac{8}{3} C_F \Nf \frac{d_F^{abcd} d_A^{abcd}}{\NA}
+ \frac{11}{8} \zeta_4 C_A^5
+ \frac{41}{2} \zeta_4 C_A^3 C_F T_F \Nf
+ \frac{1093}{2} \zeta_3^2 C_A^3 C_F T_F \Nf
\right. \nonumber \\
&& \left. ~~~~
+ \frac{1988}{9} C_A \Nf^2 \frac{d_F^{abcd} d_F^{abcd}}{\NA}
+ \frac{3655}{16} \zeta_3^2 T_F \Nf \frac{d_A^{abcd} d_A^{abcd}}{\NA}
+ \frac{6320}{3} \zeta_5 T_F \Nf^2 \frac{d_F^{abcd} d_A^{abcd}}{\NA}
\right. \nonumber \\
&& \left. ~~~~
+ \frac{13504}{27} \zeta_3 C_A^2 T_F^3 \Nf^3
+ \frac{14605}{3} \zeta_5 C_A^2 C_F^2 T_F \Nf
+ \frac{40651}{6} \zeta_3 C_A^2 C_F T_F^2 \Nf^2
\right. \nonumber \\
&& \left. ~~~~
+ \frac{42707}{6} \zeta_3 C_A \Nf \frac{d_F^{abcd} d_A^{abcd}}{\NA}
+ \frac{123395}{72} \zeta_5 C_A^3 T_F^2 \Nf^2
+ \frac{147841}{768} C_A \frac{d_A^{abcd} d_A^{abcd}}{\NA}
\right. \nonumber \\
&& \left. ~~~~
+ \frac{152269}{64} \zeta_3 T_F \Nf \frac{d_A^{abcd} d_A^{abcd}}{\NA}
+ \frac{313051}{162} C_A C_F T_F^3 \Nf^3
+ \frac{2230991}{768} \zeta_7 C_A^4 T_F \Nf
\right. \nonumber \\
&& \left. ~~~~
+ \frac{6298537}{8192} \zeta_7 C_A \frac{d_A^{abcd} d_A^{abcd}}{\NA}
+ \frac{11508905}{1536} \zeta_5 C_A \frac{d_A^{abcd} d_A^{abcd}}{\NA}
\right. \nonumber \\
&& \left. ~~~~
+ \frac{16814735}{864} C_A^3 C_F T_F \Nf
+ \frac{37229609}{8748} C_A^2 T_F^3 \Nf^3
+ \frac{189347125}{24576} \zeta_5 C_A^5
\right. \nonumber \\
&& \left. ~~~~
+ \frac{1765579181}{221184} \zeta_3 C_A^5
- 1680 \zeta_7 C_A C_F^3 T_F \Nf
- 1240 \zeta_3 C_A C_F T_F^3 \Nf^3
\right. \nonumber \\
&& \left. ~~~~
- 470 \zeta_5 C_F \Nf \frac{d_F^{abcd} d_A^{abcd}}{\NA}
- 384 \zeta_3^2 C_A C_F^2 T_F^2 \Nf^2
- 316 \zeta_3 C_A \Nf^2 \frac{d_F^{abcd} d_F^{abcd}}{\NA}
\right. \nonumber \\
&& \left. ~~~~
- 294 \zeta_7 C_A^3 T_F^2 \Nf^2
- 128 \zeta_3^2 C_A \Nf^2 \frac{d_F^{abcd} d_F^{abcd}}{\NA}
- 66 \zeta_4 C_A \frac{d_A^{abcd} d_A^{abcd}}{\NA}
\right. \nonumber \\
&& \left. ~~~~
- 48 \zeta_4 C_A \Nf^2 \frac{d_F^{abcd} d_F^{abcd}}{\NA}
- 25 C_A C_F^3 T_F \Nf
- 14 \zeta_4 C_A^2 C_F T_F^2 \Nf^2
\right. \nonumber \\
&& \left. ~~~~
- 11 \zeta_4 C_A^2 C_F^2 T_F \Nf
- 7 \zeta_4 C_A^3 T_F^2 \Nf^2
+ 22 \zeta_4 C_A C_F^2 T_F^2 \Nf^2
\right. \nonumber \\
&& \left. ~~~~
+ 32 \zeta_3^2 T_F \Nf^2 \frac{d_F^{abcd} d_A^{abcd}}{\NA}
+ 44 \zeta_3 C_F \Nf \frac{d_F^{abcd} d_A^{abcd}}{\NA}
+ 78 \zeta_3 C_A C_F^3 T_F \Nf
\right. \nonumber \\
&& \left. ~~~~
+ 144 \zeta_3^2 C_F \Nf \frac{d_F^{abcd} d_A^{abcd}}{\NA}
+ 156 \zeta_4 C_A \Nf \frac{d_F^{abcd} d_A^{abcd}}{\NA}
+ 238 \zeta_7 C_F \Nf \frac{d_F^{abcd} d_A^{abcd}}{\NA}
\right. \nonumber \\
&& \left. ~~~~
+ 240 \zeta_5 C_A \Nf^2 \frac{d_F^{abcd} d_F^{abcd}}{\NA}
+ 392 \zeta_7 C_A \Nf \frac{d_F^{abcd} d_A^{abcd}}{\NA}
+ 416 \zeta_3^2 C_A^2 C_F T_F^2 \Nf^2
\right. \nonumber \\
&& \left. ~~~~
+ 1470 \zeta_5 C_A C_F^3 T_F \Nf
+ 1680 \zeta_7 C_A^2 C_F^2 T_F \Nf
+ 1880 \zeta_5 C_A^2 C_F T_F^2 \Nf^2
\right. \nonumber \\
&& \left. ~~~~
+ 2448 \zeta_3 C_A C_F^2 T_F^2 \Nf^2
\right] a^5 ~+~ O(a^6)
\end{eqnarray}
and
\begin{eqnarray}
\gamma_\psi^{\RIs}(a,0) &=&
\left[
\frac{25}{4} C_A C_F
- \frac{3}{2} C_F^2
- 2 C_F T_F \Nf
\right] a^2 
\nonumber \\
&&
+ \left[
8 \zeta_3 C_A C_F T_F \Nf
+ 12 \zeta_3 C_A C_F^2
- \frac{764}{9} C_A C_F T_F \Nf
- \frac{245}{8} \zeta_3 C_A^2 C_F
- \frac{121}{3} C_A C_F^2
\right. \nonumber \\
&& \left. ~~~~
+ \frac{3}{2} C_F^3
+ \frac{14}{3} C_F^2 T_F \Nf
+ \frac{104}{9} C_F T_F^2 \Nf^2
+ \frac{19979}{144} C_A^2 C_F
\right] a^3 
\nonumber \\
&&
+ \left[
848 \zeta_3 C_A C_F^3
- \frac{307915}{192} \zeta_3 C_A^3 C_F
- \frac{81691}{72} C_A^2 C_F^2
- \frac{28382}{9} C_A^2 C_F T_F \Nf
\right. \nonumber \\
&& \left. ~~~~
- \frac{2000}{27} C_F T_F^3 \Nf^3
- \frac{1027}{8} C_F^4
- \frac{995}{18} C_A C_F^2 T_F \Nf
- \frac{855}{8} \zeta_5 \frac{d_F^{abcd} d_A^{abcd}}{\NF}
\right. \nonumber \\
&& \left. ~~~~
- \frac{112}{3} \zeta_3 C_A C_F T_F^2 \Nf^2
+ \frac{164}{3} C_F^3 T_F \Nf
+ \frac{547}{2} \zeta_3 C_A^2 C_F^2
+ \frac{1030}{9} C_F^2 T_F^2 \Nf^2
\right. \nonumber \\
&& \left. ~~~~
+ \frac{1082}{3} C_A C_F^3
+ \frac{1113}{8} \zeta_3 \frac{d_F^{abcd} d_A^{abcd}}{\NF}
+ \frac{2257}{3} \zeta_3 C_A^2 C_F T_F \Nf
+ \frac{7915}{9} C_A C_F T_F^2 \Nf^2
\right. \nonumber \\
&& \left. ~~~~
+ \frac{10375}{32} \zeta_5 C_A^3 C_F
+ \frac{761141}{216} C_A^3 C_F
- 1440 \zeta_5 C_A C_F^3
- 400 \zeta_3 C_F^4
\right. \nonumber \\
&& \left. ~~~~
- 245 \zeta_5 C_A^2 C_F T_F \Nf
- 96 \zeta_3 C_F^2 T_F^2 \Nf^2
- 67 \frac{d_F^{abcd} d_A^{abcd}}{\NF}
- 64 \zeta_3 C_F^3 T_F \Nf
\right. \nonumber \\
&& \left. ~~~~
+ 24 \zeta_3 C_A C_F^2 T_F \Nf
+ 128 \Nf \frac{d_F^{abcd}d_F^{abcd}}{\NF}
+ 240 \zeta_5 C_A C_F^2 T_F \Nf
+ 565 \zeta_5 C_A^2 C_F^2
\right. \nonumber \\
&& \left. ~~~~
+ 640 \zeta_5 C_F^4
\right] a^4 
\nonumber \\
&&
+ \left[
\frac{17300466253}{186624} C_A^4 C_F
- \frac{2511034225}{23328} C_A^3 C_F T_F \Nf
- \frac{685190527}{36864} \zeta_7 C_A^4 C_F
\right. \nonumber \\
&& \left. ~~~~
- \frac{135324833}{3888} C_A^3 C_F^2
- \frac{121549691}{2304} \zeta_3 C_A^4 C_F
- \frac{5756552}{729} C_A C_F T_F^3 \Nf^3
\right. \nonumber \\
&& \left. ~~~~
- \frac{3366343}{432} C_A^2 C_F^2 T_F \Nf
- \frac{2984107}{16} \zeta_7 C_A^2 C_F^3
- \frac{2272097}{48} \zeta_3 C_A^3 C_F^2
\right. \nonumber \\
&& \left. ~~~~
- \frac{1669235}{96} \zeta_3 C_A \frac{d_F^{abcd} d_A^{abcd}}{\NF}
- \frac{1458845}{384} \zeta_5 C_F \frac{d_A^{abcd} d_A^{abcd}}{\NA}
- \frac{787558}{9} \zeta_3 C_A C_F^4
\right. \nonumber \\
&& \left. ~~~~
- \frac{781985}{36} \zeta_5 C_A^2 C_F^3
- \frac{654439}{144} C_A \frac{d_F^{abcd} d_A^{abcd}}{\NF}
- \frac{543935}{24} \zeta_5 C_A^3 C_F T_F \Nf
\right. \nonumber \\
&& \left. ~~~~
- \frac{536720}{243} C_F^2 T_F^3 \Nf^3
- \frac{377000}{9} \zeta_5 C_A C_F^4
- \frac{238324}{9} \zeta_3 C_A C_F^3 T_F \Nf
\right. \nonumber \\
&& \left. ~~~~
- \frac{193093}{18} C_A C_F^4
- \frac{125447}{8} \zeta_7 C_F \frac{d_F^{abcd} d_A^{abcd}}{\NF}
- \frac{110513}{24} \zeta_3^2 C_A^3 C_F^2
\right. \nonumber \\
&& \left. ~~~~
- \frac{99622}{9} \zeta_3 C_A^2 C_F T_F^2 \Nf^2
- \frac{91085}{9} \zeta_5 C_A^2 C_F^2 T_F \Nf
- \frac{69727}{18} C_A C_F^3 T_F \Nf
\right. \nonumber \\
&& \left. ~~~~
- \frac{51928}{9} \zeta_3 C_A C_F^2 T_F^2 \Nf^2
- \frac{41909}{16} \zeta_7 C_A^3 C_F T_F \Nf
- \frac{30112}{9} T_F \Nf^2 \frac{d_F^{abcd}d_F^{abcd}}{\NF}
\right. \nonumber \\
&& \left. ~~~~
- \frac{21080}{3} \zeta_5 T_F \Nf \frac{d_F^{abcd} d_A^{abcd}}{\NF}
- \frac{20234}{27} C_F^3 T_F^2 \Nf^2
- \frac{9206}{3} \zeta_3^2 C_A^2 C_F^2 T_F \Nf
\right. \nonumber \\
&& \left. ~~~~
- \frac{7040}{27} \zeta_5 C_A C_F T_F^3 \Nf^3
- \frac{5984}{3} C_F \Nf \frac{d_F^{abcd}d_F^{abcd}}{\NF}
- \frac{5600}{3} \zeta_5 C_A C_F^2 T_F^2 \Nf^2
\right. \nonumber \\
&& \left. ~~~~
- \frac{1985}{24} C_F \frac{d_A^{abcd} d_A^{abcd}}{\NA}
- \frac{1472}{3} \zeta_3^2 C_A^2 C_F T_F^2 \Nf^2
- \frac{320}{9} \zeta_5 C_F^4 T_F \Nf
\right. \nonumber \\
&& \left. ~~~~
+ \frac{113}{6} C_F \frac{d_F^{abcd} d_A^{abcd}}{\NF}
+ \frac{1911}{2} \zeta_7 C_A^2 C_F T_F^2 \Nf^2
+ \frac{2560}{9} \zeta_5 C_F^3 T_F^2 \Nf^2
\right. \nonumber \\
&& \left. ~~~~
+ \frac{2816}{3} \zeta_3^2 C_A C_F^2 T_F^2 \Nf^2
+ \frac{3577}{64} \zeta_3^2 C_F \frac{d_A^{abcd} d_A^{abcd}}{\NA}
+ \frac{3584}{9} \zeta_3 C_A C_F T_F^3 \Nf^3
\right. \nonumber \\
&& \left. ~~~~
+ \frac{3616}{9} \zeta_3 C_F^3 T_F^2 \Nf^2
+ \frac{4041}{4} \zeta_3^2 C_A^3 C_F T_F \Nf
+ \frac{4977}{8} C_F^5
+ \frac{13568}{9} \zeta_3 C_F^2 T_F^3 \Nf^3
\right. \nonumber \\
&& \left. ~~~~
+ \frac{17120}{9} T_F \Nf \frac{d_F^{abcd} d_A^{abcd}}{\NF}
+ \frac{21760}{3} \zeta_5 T_F \Nf^2 \frac{d_F^{abcd}d_F^{abcd}}{\NF}
+ \frac{22673}{2} \zeta_3^2 C_A^2 C_F^3
\right. \nonumber \\
&& \left. ~~~~
+ \frac{23632}{3} \zeta_7 C_A \Nf \frac{d_F^{abcd}d_F^{abcd}}{\NF}
+ \frac{46165}{6} \zeta_7 T_F \Nf \frac{d_F^{abcd} d_A^{abcd}}{\NF}
+ \frac{48101}{18} C_F^4 T_F \Nf
\right. \nonumber \\
&& \left. ~~~~
+ \frac{51943}{8} \zeta_3^2 C_A \frac{d_F^{abcd} d_A^{abcd}}{\NF}
+ \frac{52550}{9} \zeta_5 C_A^2 C_F T_F^2 \Nf^2
+ \frac{53116}{3} \zeta_7 C_A^2 C_F^2 T_F \Nf
\right. \nonumber \\
&& \left. ~~~~
+ \frac{76432}{3} \zeta_3 C_A \Nf \frac{d_F^{abcd}d_F^{abcd}}{\NF}
+ \frac{96032}{9} C_A \Nf \frac{d_F^{abcd}d_F^{abcd}}{\NF}
+ \frac{115832}{9} \zeta_3 C_F^4 T_F \Nf
\right. \nonumber \\
&& \left. ~~~~
+ \frac{135731}{192} \zeta_3 C_F \frac{d_A^{abcd} d_A^{abcd}}{\NA}
+ \frac{153221}{24} \zeta_3 C_A^2 C_F^2 T_F \Nf
+ \frac{198800}{9} \zeta_5 C_A C_F^3 T_F \Nf
\right. \nonumber \\
&& \left. ~~~~
+ \frac{346240}{729} C_F T_F^4 \Nf^4
+ \frac{781753}{192} \zeta_7 C_F \frac{d_A^{abcd} d_A^{abcd}}{\NA}
+ \frac{1020625}{384} \zeta_3^2 C_A^4 C_F
\right. \nonumber \\
&& \left. ~~~~
+ \frac{1097000}{81} C_A C_F^2 T_F^2 \Nf^2
+ \frac{1155685}{96} \zeta_5 C_A \frac{d_F^{abcd} d_A^{abcd}}{\NF}
+ \frac{2432579}{18} \zeta_3 C_A^2 C_F^3
\right. \nonumber \\
&& \left. ~~~~
+ \frac{2435545}{768} \zeta_7 C_A \frac{d_F^{abcd} d_A^{abcd}}{\NF}
+ \frac{5525989}{216} C_A^2 C_F^3
+ \frac{6642965}{288} \zeta_5 C_A^3 C_F^2
\right. \nonumber \\
&& \left. ~~~~
+ \frac{13984033}{192} \zeta_7 C_A^3 C_F^2
+ \frac{14900669}{288} \zeta_3 C_A^3 C_F T_F \Nf
+ \frac{21901003}{486} C_A^2 C_F T_F^2 \Nf^2
\right. \nonumber \\
&& \left. ~~~~
+ \frac{686761615}{27648} \zeta_5 C_A^4 C_F
- 47628 \zeta_7 C_F^5
- 34400 \zeta_5 C_A \Nf \frac{d_F^{abcd}d_F^{abcd}}{\NF}
\right. \nonumber \\
&& \left. ~~~~
- 18816 \zeta_7 C_F^4 T_F \Nf
- 12096 \zeta_3 C_F \Nf \frac{d_F^{abcd}d_F^{abcd}}{\NF}
- 9384 \zeta_3^2 C_A C_F^4
\right. \nonumber \\
&& \left. ~~~~
- 8680 \zeta_7 C_F \Nf \frac{d_F^{abcd}d_F^{abcd}}{\NF}
- 7952 \zeta_3^2 C_A \Nf \frac{d_F^{abcd}d_F^{abcd}}{\NF}
- 5760 \zeta_3 T_F \Nf^2 \frac{d_F^{abcd}d_F^{abcd}}{\NF}
\right. \nonumber \\
&& \left. ~~~~
- 4884 \zeta_3^2 C_F \frac{d_F^{abcd} d_A^{abcd}}{\NF}
- 4704 \zeta_7 C_A C_F^2 T_F^2 \Nf^2
- 1027 \zeta_3 T_F \Nf \frac{d_F^{abcd} d_A^{abcd}}{\NF}
\right. \nonumber \\
&& \left. ~~~~
- 896 \zeta_3^2 C_A C_F^3 T_F \Nf
- 776 \zeta_3^2 T_F \Nf \frac{d_F^{abcd} d_A^{abcd}}{\NF}
+ 768 \zeta_3^2 C_F^4 T_F \Nf
\right. \nonumber \\
&& \left. ~~~~
+ 1015 \zeta_5 C_F \frac{d_F^{abcd} d_A^{abcd}}{\NF}
+ 2048 \zeta_3^2 T_F \Nf^2 \frac{d_F^{abcd}d_F^{abcd}}{\NF}
+ 2496 \zeta_3^2 C_F^5
\right. \nonumber \\
&& \left. ~~~~
+ 3648 \zeta_3^2 C_F \Nf \frac{d_F^{abcd}d_F^{abcd}}{\NF}
+ 13216 \zeta_7 C_A C_F^3 T_F \Nf
+ 16000 \zeta_3 C_F^5
\right. \nonumber \\
&& \left. ~~~~
+ 17554 \zeta_3 C_F \frac{d_F^{abcd} d_A^{abcd}}{\NF}
+ 18080 \zeta_5 C_F \Nf \frac{d_F^{abcd}d_F^{abcd}}{\NF}
+ 22600 \zeta_5 C_F^5
\right. \nonumber \\
&& \left. ~~~~
+ 175721 \zeta_7 C_A C_F^4
\right] a^5 
~+~ O(a^6) 
\end{eqnarray}
where $\NA$ is the dimension of the adjoint representation. Similarly the quark
mass anomalous dimension is
\begin{eqnarray}
\gamma_m^{\RIs}(a,0) &=& -~ 3 C_F a
+ \left[
\frac{26}{3} C_F T_F \Nf
- \frac{185}{6} C_A C_F
- \frac{3}{2} C_F^2
\right] a^2
\nonumber \\
&&
+ \left[
\frac{7870}{27} C_A C_F T_F \Nf
- \frac{29357}{54} C_A^2 C_F
- \frac{856}{27} C_F T_F^2 \Nf^2
- \frac{129}{2} C_F^3
- 88 \zeta_3 C_A C_F^2
\right. \nonumber \\
&& \left. ~~~~
- 16 \zeta_3 C_F^2 T_F \Nf
- 9 C_A C_F^2
+ 77 C_F^2 T_F \Nf
+ 132 \zeta_3 C_A^2 C_F
\right] a^3
\nonumber \\
&&
+ \left[
\frac{1261}{8} C_F^4
- \frac{46284559}{3888} C_A^3 C_F
- \frac{172912}{81} C_A C_F T_F^2 \Nf^2
- \frac{23104}{27} C_F^2 T_F^2 \Nf^2
\right. \nonumber \\
&& \left. ~~~~
- \frac{13102}{3} C_A C_F^3
- \frac{10987}{6} \zeta_3 C_A^2 C_F T_F \Nf
- \frac{8566}{3} \zeta_3 C_A^2 C_F^2
+ \frac{224}{3} \zeta_3 C_A C_F T_F^2 \Nf^2
\right. \nonumber \\
&& \left. ~~~~
+ \frac{520}{3} \zeta_3 C_A C_F^2 T_F \Nf
+ \frac{992}{3} \zeta_3 C_F^2 T_F^2 \Nf^2
+ \frac{3760}{3} C_F^3 T_F \Nf
+ \frac{32048}{243} C_F T_F^3 \Nf^3
\right. \nonumber \\
&& \left. ~~~~
+ \frac{95881}{72} C_A^2 C_F^2
+ \frac{120385}{24} \zeta_3 C_A^3 C_F
+ \frac{145717}{54} C_A C_F^2 T_F \Nf
\right. \nonumber \\
&& \left. ~~~~
+ \frac{3050747}{324} C_A^2 C_F T_F \Nf
- 496 \zeta_3 C_F^3 T_F \Nf
- 240 \zeta_3 \frac{d_F^{abcd} d_A^{abcd}}{N_F}
- 220 \zeta_5 C_A^2 C_F^2
\right. \nonumber \\
&& \left. ~~~~
- 220 \zeta_5 C_A^3 C_F
- 206 \zeta_3 C_A C_F^3
- 160 \zeta_5 C_A C_F^2 T_F \Nf
- 160 \zeta_5 C_A^2 C_F T_F \Nf
\right. \nonumber \\
&& \left. ~~~~
- 64 \Nf \frac{d_F^{abcd} d_F^{abcd}}{N_F}
+ 32 \frac{d_F^{abcd} d_A^{abcd}}{N_F}
+ 336 \zeta_3 C_F^4
+ 480 \zeta_3 \Nf \frac{d_F^{abcd} d_F^{abcd}}{N_F}
\right. \nonumber \\
&& \left. ~~~~
+ 1320 \zeta_5 C_A C_F^3
\right] a^4
\nonumber \\
&&
+ \left[
\frac{11}{2} \zeta_4 C_A^4 C_F
- \frac{11072937943}{34992} C_A^4 C_F
- \frac{84487076}{729} C_A^2 C_F T_F^2 \Nf^2
\right. \nonumber \\
&& \left. ~~~~
- \frac{53524919}{216} C_A^2 C_F^3
- \frac{1523584}{2187} C_F T_F^4 \Nf^4
- \frac{1445111}{27} \zeta_3 C_A^3 C_F^2
\right. \nonumber \\
&& \left. ~~~~
- \frac{1441292}{27} C_A C_F^2 T_F^2 \Nf^2
- \frac{1186031}{8} \zeta_7 C_A^3 C_F^2
- \frac{1151695}{54} C_F^3 T_F^2 \Nf^2
\right. \nonumber \\
&& \left. ~~~~
- \frac{655760}{9} \zeta_3 C_A^2 C_F^3
- \frac{627715}{18} \zeta_5 C_A^3 C_F^2
- \frac{469717}{4} \zeta_3 C_A^3 C_F T_F \Nf
\right. \nonumber \\
&& \left. ~~~~
- \frac{256531}{12} \zeta_3^2 C_A^4 C_F
- \frac{255248}{3} \zeta_7 C_A C_F^3 T_F \Nf
- \frac{217060}{9} \zeta_5 C_A^3 C_F T_F \Nf
\right. \nonumber \\
&& \left. ~~~~
- \frac{200500}{9} \zeta_5 C_A C_F^3 T_F \Nf
- \frac{198121}{4} \zeta_7 C_A \frac{d_F^{abcd} d_A^{abcd}}{N_F}
- \frac{150304}{3} \zeta_3^2 C_A^2 C_F^3
\right. \nonumber \\
&& \left. ~~~~
- \frac{110719}{6} \zeta_3 C_A \frac{d_F^{abcd} d_A^{abcd}}{N_F}
- \frac{106016}{27} \zeta_3 C_F^2 T_F^3 \Nf^3
- \frac{93317}{48} \zeta_7 C_A^3 C_F T_F \Nf
\right. \nonumber \\
&& \left. ~~~~
- \frac{69472}{9} C_A \Nf \frac{d_F^{abcd} d_F^{abcd}}{N_F}
- \frac{66344}{3} \zeta_3 C_F^4 T_F \Nf
- \frac{50995}{8} C_F^5
\right. \nonumber \\
&& \left. ~~~~
- \frac{41505}{2} \zeta_5 C_A \frac{d_F^{abcd} d_A^{abcd}}{N_F}
- \frac{39320}{3} \zeta_5 C_F^4 T_F \Nf
- \frac{21760}{27} \zeta_5 C_A C_F T_F^3 \Nf^3
\right. \nonumber \\
&& \left. ~~~~
- \frac{17312}{27} \zeta_3 C_A C_F T_F^3 \Nf^3
- \frac{16160}{9} \zeta_5 C_A C_F^2 T_F^2 \Nf^2
- \frac{10336}{9} T_F \Nf \frac{d_F^{abcd} d_A^{abcd}}{N_F}
\right. \nonumber \\
&& \left. ~~~~
- \frac{7804}{3} \zeta_3 C_F \frac{d_A^{abcd} d_A^{abcd}}{\NA}
- \frac{7040}{9} \zeta_5 C_F^3 T_F^2 \Nf^2
- \frac{6190}{3} \zeta_5 T_F \Nf \frac{d_F^{abcd} d_A^{abcd}}{N_F}
\right. \nonumber \\
&& \left. ~~~~
- \frac{2902}{9} C_F \frac{d_A^{abcd} d_A^{abcd}}{\NA}
- \frac{2048}{3} \zeta_3 C_F \Nf^2 \frac{d_F^{abcd} d_F^{abcd}}{\NA}
- \frac{2048}{9} C_F \Nf \frac{d_F^{abcd} d_A^{abcd}}{\NA}
\right. \nonumber \\
&& \left. ~~~~
+ \frac{512}{3} \zeta_3^2 C_F^3 T_F^2 \Nf^2
+ \frac{512}{3} \zeta_3^2 C_A^2 C_F T_F^2 \Nf^2
+ \frac{1280}{3} \zeta_5 C_F^2 T_F^3 \Nf^3
\right. \nonumber \\
&& \left. ~~~~
+ \frac{2031}{2} C_F^4 T_F \Nf
+ \frac{2816}{9} C_F \Nf^2 \frac{d_F^{abcd} d_F^{abcd}}{\NA}
+ \frac{6656}{3} \zeta_3 C_F \Nf \frac{d_F^{abcd} d_A^{abcd}}{\NA}
\right. \nonumber \\
&& \left. ~~~~
+ \frac{9241}{3} \zeta_3^2 C_A^3 C_F T_F \Nf
+ \frac{15640}{3} \zeta_5 C_F \frac{d_A^{abcd} d_A^{abcd}}{\NA}
+ \frac{22928}{9} T_F \Nf^2 \frac{d_F^{abcd} d_F^{abcd}}{N_F}
\right. \nonumber \\
&& \left. ~~~~
+ \frac{26200}{3} \zeta_5 C_A^2 C_F T_F^2 \Nf^2
+ \frac{35777}{9} \zeta_3 C_A C_F^3 T_F \Nf
+ \frac{37916}{3} \zeta_3 C_A C_F^2 T_F^2 \Nf^2
\right. \nonumber \\
&& \left. ~~~~
+ \frac{40576}{3} \zeta_3^2 C_A C_F^3 T_F \Nf
+ \frac{44480}{3} \zeta_5 C_A \Nf \frac{d_F^{abcd} d_F^{abcd}}{N_F}
+ \frac{54889}{2} C_A C_F^4
\right. \nonumber \\
&& \left. ~~~~
+ \frac{76349}{2} \zeta_7 C_A^2 C_F^2 T_F \Nf
+ \frac{90464}{3} \zeta_3 C_A \Nf \frac{d_F^{abcd} d_F^{abcd}}{N_F}
+ \frac{96848}{9} C_A \frac{d_F^{abcd} d_A^{abcd}}{N_F}
\right. \nonumber \\
&& \left. ~~~~
+ \frac{102650}{9} \zeta_5 C_A^2 C_F^2 T_F \Nf
+ \frac{107296}{9} \zeta_3 C_F^3 T_F^2 \Nf^2
+ \frac{163414}{3} \zeta_3 C_A C_F^4
\right. \nonumber \\
&& \left. ~~~~
+ \frac{168400}{3} \zeta_5 C_A C_F^4
+ \frac{176281}{18} \zeta_3 C_A^2 C_F^2 T_F \Nf
+ \frac{190282}{9} \zeta_3 C_A^2 C_F T_F^2 \Nf^2
\right. \nonumber \\
&& \left. ~~~~
+ \frac{380960}{27} \zeta_5 C_A^4 C_F
+ \frac{552820}{9} \zeta_5 C_A^2 C_F^3
+ \frac{757372}{3} \zeta_7 C_A^2 C_F^3
\right. \nonumber \\
&& \left. ~~~~
+ \frac{1878562}{243} C_F^2 T_F^3 \Nf^3
+ \frac{2389877}{18} C_A C_F^3 T_F \Nf
+ \frac{5635399}{192} \zeta_7 C_A^4 C_F
\right. \nonumber \\
&& \left. ~~~~
+ \frac{30274283}{648} C_A^2 C_F^2 T_F \Nf
+ \frac{34700285}{2187} C_A C_F T_F^3 \Nf^3
+ \frac{72189403}{432} \zeta_3 C_A^4 C_F
\right. \nonumber \\
&& \left. ~~~~
+ \frac{231306035}{1944} C_A^3 C_F^2
+ \frac{1441558963}{4374} C_A^3 C_F T_F \Nf
- 169400 \zeta_7 C_A C_F^4
\right. \nonumber \\
&& \left. ~~~~
- 25872 \zeta_7 C_A \Nf \frac{d_F^{abcd} d_F^{abcd}}{N_F}
- 11840 \zeta_5 C_F \Nf \frac{d_F^{abcd} d_F^{abcd}}{N_F}
\right. \nonumber \\
&& \left. ~~~~
- 11752 \zeta_3^2 T_F \Nf \frac{d_F^{abcd} d_A^{abcd}}{N_F}
- 11376 \zeta_3 T_F \Nf^2 \frac{d_F^{abcd} d_F^{abcd}}{N_F}
- 7808 \zeta_3^2 C_F^4 T_F \Nf
\right. \nonumber \\
&& \left. ~~~~
- 6768 C_F \frac{d_F^{abcd} d_A^{abcd}}{N_F}
- 4448 \zeta_3^2 C_A^2 C_F^2 T_F \Nf
- 4224 \zeta_3^2 C_A \Nf \frac{d_F^{abcd} d_F^{abcd}}{N_F}
\right. \nonumber \\
&& \left. ~~~~
- 3872 \zeta_3^2 C_F \frac{d_A^{abcd} d_A^{abcd}}{\NA}
- 3840 \zeta_3 C_F \frac{d_F^{abcd} d_A^{abcd}}{N_F}
- 2156 \zeta_7 C_A^2 C_F T_F^2 \Nf^2
\right. \nonumber \\
&& \left. ~~~~
- 2080 \zeta_5 C_F^5
- 1152 \zeta_3^2 C_A C_F^2 T_F^2 \Nf^2
- 848 \zeta_3 C_F^5
- 320 \zeta_5 T_F \Nf^2 \frac{d_F^{abcd} d_F^{abcd}}{N_F}
\right. \nonumber \\
&& \left. ~~~~
- 264 \zeta_4 C_F \frac{d_A^{abcd} d_A^{abcd}}{\NA}
- 192 \zeta_4 T_F \Nf^2 \frac{d_F^{abcd} d_F^{abcd}}{N_F}
- 56 \zeta_4 C_A C_F^2 T_F^2 \Nf^2
\right. \nonumber \\
&& \left. ~~~~
- 51 \zeta_4 C_A^3 C_F T_F \Nf
- 44 \zeta_4 C_A C_F^3 T_F \Nf
- 28 \zeta_4 C_A^2 C_F T_F^2 \Nf^2
\right. \nonumber \\
&& \left. ~~~~
+ 82 \zeta_4 C_A^2 C_F^2 T_F \Nf
+ 88 \zeta_4 C_F^3 T_F^2 \Nf^2
+ 352 C_F \Nf \frac{d_F^{abcd} d_F^{abcd}}{N_F}
\right. \nonumber \\
&& \left. ~~~~
+ 624 \zeta_4 T_F \Nf \frac{d_F^{abcd} d_A^{abcd}}{N_F}
+ 1024 \zeta_3^2 T_F \Nf^2 \frac{d_F^{abcd} d_F^{abcd}}{N_F}
+ 1232 \zeta_7 C_F \frac{d_A^{abcd} d_A^{abcd}}{\NA}
\right. \nonumber \\
&& \left. ~~~~
+ 4704 \zeta_7 C_A C_F^2 T_F^2 \Nf^2
+ 4928 \zeta_3 C_F \Nf \frac{d_F^{abcd} d_F^{abcd}}{N_F}
+ 9408 \zeta_7 T_F \Nf^2 \frac{d_F^{abcd} d_F^{abcd}}{N_F}
\right. \nonumber \\
&& \left. ~~~~
+ 11830 \zeta_3 T_F \Nf \frac{d_F^{abcd} d_A^{abcd}}{N_F}
+ 17563 \zeta_7 T_F \Nf \frac{d_F^{abcd} d_A^{abcd}}{N_F}
+ 21472 \zeta_3^2 C_A C_F^4
\right. \nonumber \\
&& \left. ~~~~
+ 25440 \zeta_5 C_F \frac{d_F^{abcd} d_A^{abcd}}{N_F}
+ 40062 \zeta_3^2 C_A \frac{d_F^{abcd} d_A^{abcd}}{N_F}
+ 44176 \zeta_3^2 C_A^3 C_F^2
\right. \nonumber \\
&& \left. ~~~~
+ 54880 \zeta_7 C_F^4 T_F \Nf
\right] a^5 ~+~ O(a^6) ~.
\end{eqnarray}
One ingredient that was necessary in applying (\ref{anomdimmap}) to find the
above expressions was the relation between the gauge parameter in the
$\MSbar$ and $\RI$ schemes. Therefore we record that 
\begin{eqnarray}
\alpha_{\RIs} &=& \left[ 1
+ \left[
\frac{20}{9} T_F \Nf
- \frac{97}{36} C_A
- \frac{1}{2} C_A \alpha
- \frac{1}{4} C_A \alpha^2
\right] a
\right. \nonumber \\
&& \left.
+ \left[
\frac{1}{16} C_A^2 \alpha^4
- \frac{2381}{96} C_A^2
- \frac{10}{9} C_A T_F \Nf \alpha
- \frac{10}{9} C_A T_F \Nf \alpha^2
- \frac{1}{16} C_A^2 \alpha^3
+ \frac{55}{3} C_F T_F \Nf
\right. \right. \nonumber \\
&& \left. \left. ~~~
+ \frac{59}{4} C_A T_F \Nf
+ \frac{95}{144} C_A^2 \alpha^2
+ \frac{463}{288} C_A^2 \alpha
- 16 \zeta_3 C_F T_F \Nf
- 2 \zeta_3 C_A^2 \alpha
+ 3 \zeta_3 C_A^2
\right. \right. \nonumber \\
&& \left. \left. ~~~
+ 8 \zeta_3 C_A T_F \Nf
\right] a^2
\right. \nonumber \\
&& \left.
+ \left[
\frac{1}{16} C_A^3 \alpha^5
- \frac{10221367}{31104} C_A^3
- \frac{12071}{288} \zeta_3 C_A^3 \alpha
- \frac{10499}{243} C_A T_F^2 \Nf^2
- \frac{7402}{81} C_F T_F^2 \Nf^2
\right. \right. \nonumber \\
&& \left. \left. ~~~
- \frac{6137}{1296} C_A^2 T_F \Nf \alpha^2
- \frac{2813}{1152} C_A^3 \alpha^3
- \frac{1492}{9} \zeta_3 C_A C_F T_F \Nf
- \frac{296}{3} \zeta_3 C_F^2 T_F \Nf
\right. \right. \nonumber \\
&& \left. \left. ~~~
- \frac{286}{9} C_F^2 T_F \Nf
- \frac{256}{9} \zeta_3 C_A T_F^2 \Nf^2
- \frac{241}{24} C_A^2 T_F \Nf \alpha
- \frac{161}{96} \zeta_3 C_A^3 \alpha^2
\right. \right. \nonumber \\
&& \left. \left. ~~~
- \frac{160}{3} \zeta_5 C_A^2 T_F \Nf
- \frac{100}{81} C_A T_F^2 \Nf^2 \alpha^2
- \frac{64}{9} \zeta_3 C_A T_F^2 \Nf^2 \alpha
- \frac{55}{6} C_A C_F T_F \Nf \alpha
\right. \right. \nonumber \\
&& \left. \left. ~~~
- \frac{55}{6} C_A C_F T_F \Nf \alpha^2
- \frac{35}{192} \zeta_5 C_A^3 \alpha^4
- \frac{29}{48} C_A^3 \alpha^4
- \frac{25}{6} \zeta_3 C_A^2 T_F \Nf \alpha^2
- \frac{5}{18} C_A^2 T_F \Nf \alpha^3
\right. \right. \nonumber \\
&& \left. \left. ~~~
- \frac{5}{24} \zeta_5 C_A^3 \alpha^3
- \frac{1}{64} C_A^3 \alpha^6
+ \frac{3}{8} \zeta_4 C_A^3 \alpha
+ \frac{3}{32} \zeta_4 C_A^3 \alpha^2
+ \frac{5}{12} C_A^2 T_F \Nf \alpha^4
+ \frac{9}{32} \zeta_4 C_A^3
\right. \right. \nonumber \\
&& \left. \left. ~~~
+ \frac{13}{96} \zeta_3 C_A^3 \alpha^4
+ \frac{16}{9} C_A T_F^2 \Nf^2 \alpha
+ \frac{115}{8} \zeta_5 C_A^3 \alpha
+ \frac{149}{96} \zeta_3 C_A^3 \alpha^3
+ \frac{202}{9} \zeta_3 C_A^2 T_F \Nf \alpha
\right. \right. \nonumber \\
&& \left. \left. ~~~
+ \frac{385}{96} \zeta_5 C_A^3 \alpha^2
+ \frac{608}{9} \zeta_3 C_F T_F^2 \Nf^2
+ \frac{871}{18} \zeta_3 C_A^2 T_F \Nf
+ \frac{1549}{24} \zeta_3 C_A^3
+ \frac{7025}{192} \zeta_5 C_A^3
\right. \right. \nonumber \\
&& \left. \left. ~~~
+ \frac{13141}{1152} C_A^3 \alpha
+ \frac{30835}{10368} C_A^3 \alpha^2
+ \frac{96809}{324} C_A C_F T_F \Nf
+ \frac{1154561}{3888} C_A^2 T_F \Nf
\right. \right. \nonumber \\
&& \left. \left. ~~~
- 80 \zeta_5 C_A C_F T_F \Nf
- 9 \zeta_4 C_A^2 T_F \Nf
+ 8 \zeta_3 C_A C_F T_F \Nf \alpha
+ 8 \zeta_3 C_A C_F T_F \Nf \alpha^2
\right. \right. \nonumber \\
&& \left. \left. ~~~
+ 12 \zeta_4 C_A C_F T_F \Nf
+ 160 \zeta_5 C_F^2 T_F \Nf
\right] a^3
\right. \nonumber \\
&& \left.
+ \left[
\frac{1}{6} \zeta_3^2 C_A^3 T_F \Nf \alpha^3
- \frac{4698915983}{746496} C_A^4
- \frac{51101257}{82944} \zeta_3 C_A^4 \alpha
- \frac{44177531}{23328} C_A^2 T_F^2 \Nf^2
\right. \right. \nonumber \\
&& \left. \left. ~~~
- \frac{21636697}{186624} C_A^3 T_F \Nf \alpha
- \frac{13241219}{2916} C_A C_F T_F^2 \Nf^2
- \frac{9266411}{6144} \zeta_7 C_A^4
- \frac{7806659}{36864} \zeta_7 C_A^4 \alpha
\right. \right. \nonumber \\
&& \left. \left. ~~~
- \frac{2360029}{18432} \zeta_3^2 C_A^4
- \frac{1650287}{432} \zeta_5 C_A^3 T_F \Nf
- \frac{1100549}{41472} C_A^4 \alpha^3
- \frac{576533}{648} C_A C_F^2 T_F \Nf
\right. \right. \nonumber \\
&& \left. \left. ~~~
- \frac{515119}{55296} \zeta_5 C_A^4 \alpha^3
- \frac{485489}{2592} C_A^2 C_F T_F \Nf \alpha
- \frac{324935}{162} \zeta_3 C_A^2 C_F T_F \Nf
\right. \right. \nonumber \\
&& \left. \left. ~~~
- \frac{311113}{768} \zeta_3 \frac{d_A^{abcd} d_A^{abcd}}{\NA}
- \frac{310055}{110592} \zeta_5 C_A^4 \alpha^4
- \frac{298517}{3456} C_A^3 T_F \Nf \alpha^2
- \frac{238325}{4096} \zeta_6 C_A^4
\right. \right. \nonumber \\
&& \left. \left. ~~~
- \frac{232351}{1296} \zeta_3 C_A^3 T_F \Nf
- \frac{226975}{512} \zeta_5 \frac{d_A^{abcd} d_A^{abcd}}{\NA}
- \frac{193639}{4608} \zeta_3 C_A^4 \alpha^2
\right. \right. \nonumber \\
&& \left. \left. ~~~
- \frac{192157}{54} \zeta_3 C_A C_F^2 T_F \Nf
- \frac{156817}{162} \zeta_3 C_A^2 T_F^2 \Nf^2
- \frac{147553}{6144} \zeta_7 C_A^4 \alpha^2
\right. \right. \nonumber \\
&& \left. \left. ~~~
- \frac{99757}{128} \zeta_7 \frac{d_A^{abcd} d_A^{abcd}}{\NA}
- \frac{82181}{648} C_A^2 C_F T_F \Nf \alpha^2
- \frac{51175}{6144} \zeta_6 C_A^4 \alpha
\right. \right. \nonumber \\
&& \left. \left. ~~~
- \frac{50925}{512} \zeta_6 \frac{d_A^{abcd} d_A^{abcd}}{\NA}
- \frac{36166}{9} \zeta_5 C_A^2 C_F T_F \Nf
- \frac{33515}{192} \zeta_3 \frac{d_A^{abcd} d_A^{abcd}}{\NA} \alpha
\right. \right. \nonumber \\
&& \left. \left. ~~~
- \frac{25888}{81} \zeta_3 C_F T_F^3 \Nf^3
- \frac{22703}{144} \zeta_5 C_A^3 T_F \Nf \alpha
- \frac{17449}{128} \zeta_3^2 \frac{d_A^{abcd} d_A^{abcd}}{\NA} \alpha
\right. \right. \nonumber \\
&& \left. \left. ~~~
- \frac{16000}{9} \zeta_5 C_F^2 T_F^2 \Nf^2
- \frac{14803}{2048} C_A^4 \alpha^4
- \frac{8369}{288} \zeta_3 C_A^3 T_F \Nf \alpha^2
- \frac{7505}{162} C_F^2 T_F^2 \Nf^2
\right. \right. \nonumber \\
&& \left. \left. ~~~
- \frac{5452}{27} \zeta_3 C_A^2 T_F^2 \Nf^2 \alpha
- \frac{5350}{3} \zeta_3^2 \Nf \frac{d_F^{abcd} d_A^{abcd}}{\NA}
- \frac{5195}{128} \zeta_5 \frac{d_A^{abcd} d_A^{abcd}}{\NA} \alpha^2
\right. \right. \nonumber \\
&& \left. \left. ~~~
- \frac{4751}{1296} C_A^3 T_F \Nf \alpha^3
- \frac{4565}{9216} \zeta_5 C_A^4 \alpha^5
- \frac{4544}{27} \Nf \frac{d_F^{abcd} d_A^{abcd}}{\NA}
- \frac{3997}{384} \zeta_7 \frac{d_A^{abcd} d_A^{abcd}}{\NA} \alpha
\right. \right. \nonumber \\
&& \left. \left. ~~~
- \frac{3255}{256} \zeta_5 \frac{d_A^{abcd} d_A^{abcd}}{\NA} \alpha^3
- \frac{2800}{3} \zeta_7 C_A^2 C_F T_F \Nf
- \frac{2767}{32} \zeta_4 C_A^3 T_F \Nf
\right. \right. \nonumber \\
&& \left. \left. ~~~
- \frac{2375}{256} \zeta_6 \frac{d_A^{abcd} d_A^{abcd}}{\NA} \alpha
- \frac{1501}{64} \zeta_3^2 \frac{d_A^{abcd} d_A^{abcd}}{\NA} \alpha^2
- \frac{1280}{9} \zeta_5 C_F T_F^3 \Nf^3
\right. \right. \nonumber \\
&& \left. \left. ~~~
- \frac{1213}{108} \zeta_3 C_A^3 T_F \Nf \alpha^3
- \frac{1195}{768} \zeta_3 \frac{d_A^{abcd} d_A^{abcd}}{\NA} \alpha^4
- \frac{989}{32} \zeta_4 \frac{d_A^{abcd} d_A^{abcd}}{\NA}
- \frac{985}{3072} \zeta_3^2 C_A^4 \alpha^3
\right. \right. \nonumber \\
&& \left. \left. ~~~
- \frac{875}{1728} C_A^3 T_F \Nf \alpha^4
- \frac{784}{9} \zeta_3 C_A C_F T_F^2 \Nf^2 \alpha
- \frac{704}{9} \zeta_3^2 C_A^2 T_F^2 \Nf^2
- \frac{640}{27} \zeta_5 C_A T_F^3 \Nf^3
\right. \right. \nonumber \\
&& \left. \left. ~~~
- \frac{633}{128} \zeta_3^2 \frac{d_A^{abcd} d_A^{abcd}}{\NA} \alpha^3
- \frac{567}{128} \zeta_7 \frac{d_A^{abcd} d_A^{abcd}}{\NA} \alpha^4
- \frac{557}{18} \zeta_3^2 C_A^3 T_F \Nf
\right. \right. \nonumber \\
&& \left. \left. ~~~
- \frac{512}{3} \zeta_3^2 \Nf^2 \frac{d_F^{abcd} d_F^{abcd}}{\NA}
- \frac{487}{384} \frac{d_A^{abcd} d_A^{abcd}}{\NA} \alpha^2
- \frac{385}{96} \frac{d_A^{abcd} d_A^{abcd}}{\NA} \alpha
- \frac{320}{81} C_A T_F^3 \Nf^3 \alpha
\right. \right. \nonumber \\
&& \left. \left. ~~~
- \frac{189}{256} \zeta_7 C_A^4 \alpha^4
- \frac{181}{2048} \zeta_4 C_A^4 \alpha^4
- \frac{175}{216} \zeta_5 C_A^3 T_F \Nf \alpha^4
- \frac{175}{512} \zeta_6 C_A^4 \alpha^2
- \frac{157}{384} \zeta_3 C_A^4 \alpha^5
\right. \right. \nonumber \\
&& \left. \left. ~~~
- \frac{147}{2} \zeta_7 \Nf \frac{d_F^{abcd} d_A^{abcd}}{\NA} \alpha
- \frac{147}{256} \zeta_3^2 C_A^4 \alpha^2
- \frac{121}{1536} C_A^4 \alpha^5
- \frac{97}{81} C_A^2 T_F^2 \Nf^2 \alpha^3
\right. \right. \nonumber \\
&& \left. \left. ~~~
- \frac{91}{192} \zeta_7 \frac{d_A^{abcd} d_A^{abcd}}{\NA} \alpha^5
- \frac{73}{256} \zeta_4 C_A^4 \alpha^3
- \frac{55}{24} C_A^2 C_F T_F \Nf \alpha^3
- \frac{41}{6} \zeta_4 C_A^2 T_F^2 \Nf^2
\right. \right. \nonumber \\
&& \left. \left. ~~~
- \frac{31}{3} C_F^3 T_F \Nf
- \frac{13}{192} \zeta_3 C_A^4 \alpha^6
- \frac{10}{9} \zeta_5 C_A^2 T_F^2 \Nf^2 \alpha^2
- \frac{7}{256} C_A^4 \alpha^7
- \frac{5}{36} C_A^3 T_F \Nf \alpha^6
\right. \right. \nonumber \\
&& \left. \left. ~~~
- \frac{5}{64} \zeta_3 \frac{d_A^{abcd} d_A^{abcd}}{\NA} \alpha^5
- \frac{3}{2} \zeta_4 C_A^2 C_F T_F \Nf \alpha
+ \frac{1}{256} C_A^4 \alpha^8
+ \frac{3}{64} \zeta_4 \frac{d_A^{abcd} d_A^{abcd}}{\NA} \alpha^4
\right. \right. \nonumber \\
&& \left. \left. ~~~
+ \frac{3}{128} \zeta_4 \frac{d_A^{abcd} d_A^{abcd}}{\NA} \alpha^3
+ \frac{5}{6} \zeta_4 C_A^3 T_F \Nf \alpha
+ \frac{5}{6} \zeta_3^2 C_A^3 T_F \Nf \alpha^2
+ \frac{5}{12} C_A^3 T_F \Nf \alpha^5
\right. \right. \nonumber \\
&& \left. \left. ~~~
+ \frac{8}{3} \zeta_4 C_A T_F^3 \Nf^3
+ \frac{9}{8} \zeta_4 \frac{d_A^{abcd} d_A^{abcd}}{\NA} \alpha^2
+ \frac{15}{64} \zeta_3^2 \frac{d_A^{abcd} d_A^{abcd}}{\NA} \alpha^5
+ \frac{21}{512} \zeta_3^2 C_A^4 \alpha^5
\right. \right. \nonumber \\
&& \left. \left. ~~~
+ \frac{25}{27} C_A^2 T_F^2 \Nf^2 \alpha^4
+ \frac{25}{512} \zeta_6 \frac{d_A^{abcd} d_A^{abcd}}{\NA} \alpha^4
+ \frac{29}{6} \zeta_3^2 C_A^3 T_F \Nf \alpha
+ \frac{32}{9} \zeta_3 C_A^2 T_F^2 \Nf^2 \alpha^3
\right. \right. \nonumber \\
&& \left. \left. ~~~
+ \frac{35}{384} \zeta_5 \frac{d_A^{abcd} d_A^{abcd}}{\NA} \alpha^5
+ \frac{35}{384} \zeta_5 C_A^4 \alpha^6
+ \frac{41}{6} \zeta_4 C_A^3 T_F \Nf \alpha^2
+ \frac{49}{2} \zeta_7 \Nf \frac{d_F^{abcd} d_A^{abcd}}{\NA}
\right. \right. \nonumber \\
&& \left. \left. ~~~
+ \frac{50}{9} \zeta_3 C_A^2 T_F^2 \Nf^2 \alpha^2
+ \frac{55}{16} C_A^2 C_F T_F \Nf \alpha^4
+ \frac{59}{27} \zeta_3 C_A^3 T_F \Nf \alpha^4
+ \frac{88}{3} \zeta_4 C_F^2 T_F^2 \Nf^2
\right. \right. \nonumber \\
&& \left. \left. ~~~
+ \frac{143}{9} C_A C_F^2 T_F \Nf \alpha
+ \frac{143}{9} C_A C_F^2 T_F \Nf \alpha^2
+ \frac{148}{3} \zeta_3 C_A C_F^2 T_F \Nf \alpha
\right. \right. \nonumber \\
&& \left. \left. ~~~
+ \frac{148}{3} \zeta_3 C_A C_F^2 T_F \Nf \alpha^2
+ \frac{227}{36} C_A^2 T_F^2 \Nf^2 \alpha^2
+ \frac{230}{9} \zeta_5 C_A^2 T_F^2 \Nf^2 \alpha
+ \frac{245}{6} \zeta_4 C_A C_F^2 T_F \Nf
\right. \right. \nonumber \\
&& \left. \left. ~~~
+ \frac{275}{4} \zeta_6 C_A^3 T_F \Nf
+ \frac{275}{256} \zeta_6 \frac{d_A^{abcd} d_A^{abcd}}{\NA} \alpha^3
+ \frac{283}{64} \zeta_3 \frac{d_A^{abcd} d_A^{abcd}}{\NA} \alpha^3
+ \frac{329}{256} \zeta_3^2 \frac{d_A^{abcd} d_A^{abcd}}{\NA} \alpha^4
\right. \right. \nonumber \\
&& \left. \left. ~~~
+ \frac{341}{1152} C_A^4 \alpha^6
+ \frac{365}{1536} \zeta_4 C_A^4 \alpha^2
+ \frac{375}{128} \zeta_6 \frac{d_A^{abcd} d_A^{abcd}}{\NA} \alpha^2
+ \frac{413}{6} \zeta_3 C_A^2 C_F T_F \Nf \alpha^2
\right. \right. \nonumber \\
&& \left. \left. ~~~
+ \frac{475}{6144} \zeta_6 C_A^4 \alpha^3
+ \frac{539}{8} \zeta_7 C_A^3 T_F \Nf \alpha
+ \frac{575}{12288} \zeta_6 C_A^4 \alpha^4
+ \frac{595}{432} \zeta_5 C_A^3 T_F \Nf \alpha^3
\right. \right. \nonumber \\
&& \left. \left. ~~~
+ \frac{992}{3} \zeta_3 \Nf \frac{d_F^{abcd} d_A^{abcd}}{\NA} \alpha
+ \frac{1279}{6144} \zeta_3^2 C_A^4 \alpha^4
+ \frac{1280}{81} \zeta_3 C_A T_F^3 \Nf^3 \alpha
\right. \right. \nonumber \\
&& \left. \left. ~~~
+ \frac{1313}{96} \zeta_3 \frac{d_A^{abcd} d_A^{abcd}}{\NA} \alpha^2
+ \frac{1347}{128} \zeta_4 \frac{d_A^{abcd} d_A^{abcd}}{\NA} \alpha
+ \frac{1397}{48} \zeta_5 C_A^3 T_F \Nf \alpha^2
\right. \right. \nonumber \\
&& \left. \left. ~~~
+ \frac{1625}{9} \zeta_3 C_A^2 C_F T_F \Nf \alpha
+ \frac{2051}{81} C_A C_F T_F^2 \Nf^2 \alpha^2
+ \frac{2625}{64} \zeta_7 \frac{d_A^{abcd} d_A^{abcd}}{\NA} \alpha^2
\right. \right. \nonumber \\
&& \left. \left. ~~~
+ \frac{2737}{384} \zeta_7 \frac{d_A^{abcd} d_A^{abcd}}{\NA} \alpha^3
+ \frac{3319}{3456} \frac{d_A^{abcd} d_A^{abcd}}{\NA}
+ \frac{4280}{27} \zeta_3 C_A T_F^3 \Nf^3
+ \frac{4505}{1536} \zeta_5 \frac{d_A^{abcd} d_A^{abcd}}{\NA} \alpha^4
\right. \right. \nonumber \\
&& \left. \left. ~~~
+ \frac{4781}{81} C_A C_F T_F^2 \Nf^2 \alpha
+ \frac{4823}{36864} \zeta_7 C_A^4 \alpha^5
+ \frac{6008}{3} \zeta_5 C_A C_F T_F^2 \Nf^2
+ \frac{6896}{27} \Nf^2 \frac{d_F^{abcd} d_F^{abcd}}{\NA}
\right. \right. \nonumber \\
&& \left. \left. ~~~
+ \frac{8411}{1536} \zeta_4 C_A^4 \alpha
+ \frac{9471}{4} \zeta_7 C_A^3 T_F \Nf
+ \frac{15785}{9216} \zeta_7 C_A^4 \alpha^3
+ \frac{16586}{27} \zeta_5 C_A^2 T_F^2 \Nf^2
\right. \right. \nonumber \\
&& \left. \left. ~~~
+ \frac{29210}{9} \zeta_5 C_A C_F^2 T_F \Nf
+ \frac{49117}{2048} \zeta_4 C_A^4
+ \frac{49696}{27} \zeta_3 C_F^2 T_F^2 \Nf^2
+ \frac{63007}{3072} \zeta_3^2 C_A^4 \alpha
\right. \right. \nonumber \\
&& \left. \left. ~~~
+ \frac{69395}{1152} \zeta_5 C_A^4 \alpha^2
+ \frac{90341}{27648} \zeta_3 C_A^4 \alpha^4
+ \frac{113743}{1458} C_A T_F^3 \Nf^3
+ \frac{124210}{81} \zeta_3 C_A C_F T_F^2 \Nf^2
\right. \right. \nonumber \\
&& \left. \left. ~~~
+ \frac{260717}{432} \zeta_3 C_A^3 T_F \Nf \alpha
+ \frac{322195}{768} \zeta_5 \frac{d_A^{abcd} d_A^{abcd}}{\NA} \alpha
+ \frac{393026}{729} C_F T_F^3 \Nf^3
\right. \right. \nonumber \\
&& \left. \left. ~~~
+ \frac{552001}{11664} C_A^2 T_F^2 \Nf^2 \alpha
+ \frac{651137}{27648} \zeta_3 C_A^4 \alpha^3
+ \frac{753337}{768} \zeta_3^2 \frac{d_A^{abcd} d_A^{abcd}}{\NA}
+ \frac{882095}{27648} C_A^4 \alpha^2
\right. \right. \nonumber \\
&& \left. \left. ~~~
+ \frac{1847129}{6144} \zeta_5 C_A^4 \alpha
+ \frac{20572427}{2916} C_A^2 C_F T_F \Nf
+ \frac{42950657}{5832} C_A^3 T_F \Nf
\right. \right. \nonumber \\
&& \left. \left. ~~~
+ \frac{122449741}{82944} \zeta_3 C_A^4
+ \frac{129316433}{1492992} C_A^4 \alpha
+ \frac{338738527}{110592} \zeta_5 C_A^4
- 2240 \zeta_7 C_F^3 T_F \Nf
\right. \right. \nonumber \\
&& \left. \left. ~~~
- 340 \zeta_5 \Nf \frac{d_F^{abcd} d_A^{abcd}}{\NA} \alpha
- 336 \zeta_3 \Nf^2 \frac{d_F^{abcd} d_F^{abcd}}{\NA}
- 256 \zeta_3^2 C_F^2 T_F^2 \Nf^2
\right. \right. \nonumber \\
&& \left. \left. ~~~
- 150 \zeta_6 C_A C_F^2 T_F \Nf
- 80 \zeta_5 C_A C_F^2 T_F \Nf \alpha
- 80 \zeta_5 C_A C_F^2 T_F \Nf \alpha^2
\right. \right. \nonumber \\
&& \left. \left. ~~~
- 64 \zeta_4 \Nf^2 \frac{d_F^{abcd} d_F^{abcd}}{\NA}
- 60 \zeta_3^2 C_A C_F^2 T_F \Nf
- 22 \zeta_4 C_A C_F T_F^2 \Nf^2
\right. \right. \nonumber \\
&& \left. \left. ~~~
- 16 \zeta_3 C_A C_F T_F^2 \Nf^2 \alpha^2
- 9 \zeta_4 C_A^2 C_F T_F \Nf \alpha^2
- 3 \zeta_3 C_A^2 C_F T_F \Nf \alpha^4
\right. \right. \nonumber \\
&& \left. \left. ~~~
+ \zeta_4 C_A^2 T_F^2 \Nf^2 \alpha
+ 2 \zeta_3 C_A^2 C_F T_F \Nf \alpha^3
+ 16 \zeta_3^2 \Nf \frac{d_F^{abcd} d_A^{abcd}}{\NA} \alpha
\right. \right. \nonumber \\
&& \left. \left. ~~~
+ 40 \zeta_5 C_A^2 C_F T_F \Nf \alpha^2
+ 61 \zeta_4 C_A^2 C_F T_F \Nf
+ 67 \zeta_5 C_A^2 C_F T_F \Nf \alpha
\right. \right. \nonumber \\
&& \left. \left. ~~~
+ 75 \zeta_6 \Nf \frac{d_F^{abcd} d_A^{abcd}}{\NA}
+ 75 \zeta_6 C_A^2 C_F T_F \Nf
+ 104 \zeta_3 C_F^3 T_F \Nf
+ 172 \zeta_4 \Nf \frac{d_F^{abcd} d_A^{abcd}}{\NA}
\right. \right. \nonumber \\
&& \left. \left. ~~~
+ 256 \zeta_3^2 C_A C_F T_F^2 \Nf^2
+ 320 \zeta_5 \Nf^2 \frac{d_F^{abcd} d_F^{abcd}}{\NA}
+ 410 \zeta_3^2 C_A^2 C_F T_F \Nf
\right. \right. \nonumber \\
&& \left. \left. ~~~
+ 1005 \zeta_5 \Nf \frac{d_F^{abcd} d_A^{abcd}}{\NA}
+ 1960 \zeta_5 C_F^3 T_F \Nf
+ 1964 \zeta_3 \Nf \frac{d_F^{abcd} d_A^{abcd}}{\NA}
\right. \right. \nonumber \\
&& \left. \left. ~~~
+ 2240 \zeta_7 C_A C_F^2 T_F \Nf
\right] a^4 ~+~ O(a^5) \right] \alpha
\end{eqnarray}
is the full mapping in an arbitrary gauge.


\begin{thebibliography}{99}
\bibitem{1} G. Martinelli, C. Pittori, C.T. Sachrajda, M. Testa \& A.
Vladikas, Nucl. Phys. {\bf B445} (1995), 81.
\bibitem{2} E. Franco \& V. Lubicz, Nucl. Phys. {\bf B531} (1998), 641.
\bibitem{3} C. Sachrajda, PoS LATTICE2010 (2010), 018.
\bibitem{4} G. 't Hooft, Nucl. Phys. {\bf B61} (1973), 455.
\bibitem{5} W.A. Bardeen, A.J. Buras, D.W. Duke \& T. Muta, Phys. Rev.
{\bf D18} (1978), 3998.
\bibitem{6} P.A. Baikov, K.G. Chetyrkin \& J.H. K\"{u}hn, Phys. Rev. Lett.
{\bf 118} (2017), 082002.
\bibitem{7} F. Herzog, B. Ruijl, T. Ueda, J.A.M. Vermaseren \& A. Vogt, JHEP
{\bf 02} (2017), 090.
\bibitem{8} T. Luthe, A. Maier, P. Marquard \& Y. Schr\"{o}der, JHEP {\bf 03}
(2017), 020.
\bibitem{9} T. Luthe, A. Maier, P. Marquard \& Y. Schr\"{o}der, JHEP {\bf 10}
(2017), 166.
\bibitem{10} K.G. Chetyrkin, G. Falcioni, F. Herzog \& J.A.M. Vermaseren, JHEP
{\bf 10} (2017), 179.
\bibitem{11} K.G. Chetyrkin \& A. R\'{e}tey, Nucl. Phys. {\bf B583} (2000), 3.
\bibitem{12} K.G. Chetyrkin \& A. R\'{e}tey, hep-ph/0007088.
\bibitem{13} J.A. Gracey, Nucl. Phys. {\bf B662} (2003), 247. 
\bibitem{14} F. He, Y.-J. Bi, T. Draper, K.-F. Liu, Z. Liu \& Y.-B. Tang,
arXiv:2204.09246 [hep-lat].
\bibitem{15} J. De Blas, Y. Du, C. Grojean, J. Gu, V. Miralles, M.E. Peskin, J.
Tian, M. Vos \& E. Vryonidou, arXiv:2206.08326 [hep-ph].
\bibitem{16} R. Tsuji, N. Tsukamoto, Y. Aoki, K.-I. Ishikawa, Y. Kuramashi, S.
Sasaki, E. Shintani \& T. Yamazaki, Phys. Rev. {\bf D106} (2022), 094505.
\bibitem{17} W.E. Caswell, Phys. Rev. Lett. {\bf 33} (1974), 244.
\bibitem{18} T. Banks \& A. Zaks, Nucl. Phys. {\bf B196} (1982), 189.
\bibitem{19} T.A. Ryttov, Phys. Rev. {\bf D89} (2014), 016013.
\bibitem{20} T.A. Ryttov, Phys. Rev. {\bf D90} (2014), 056007; Phys. Rev.
{\bf D91} (2015), 039906(E).
\bibitem{21} T.A. Ryttov \& R. Shrock, Phys. Rev. {\bf D94} (2016), 105015.
\bibitem{22} paper in preparation.
\bibitem{23} A. Cheng, A. Hasenfratz, Y. Liu, G. Petropoulos \& D. Schaich,
Phys. Rev. {\bf D90} (2014), 014509.
\bibitem{24} M.P. Lombardo, K. Miura, T.J. Nunes da Silva \& E. Pallante, JHEP
{\bf 1412} (2014), 183.
\bibitem{25} W. Celmaster \& R.J. Gonsalves, Phys. Rev. Lett. {\bf 42} (1979),
1435.
\bibitem{26} W. Celmaster \& R.J. Gonsalves, Phys. Rev. {\bf D20} (1979),
1420.
\bibitem{27} S.G. Gorishny, S.A. Larin, L.R. Surguladze \& F.K. Tkachov,
Comput. Phys. Commun. {\bf 55} (1989), 381.
\bibitem{28} S.A. Larin, F.V. Tkachov \& J.A.M. Vermaseren, The Form version
of Mincer, NIKHEF-H-91-18.
\bibitem{29} T. van Ritbergen, J.A.M. Vermaseren \& S.A. Larin, Phys. Lett.
{\bf B400} (1997), 379.
\bibitem{30} M. Czakon, Nucl. Phys. {\bf B710} (2005), 485.
\bibitem{31} K.G. Chetyrkin, Phys. Lett. {\bf B404} (1997), 161.
\bibitem{32} J.A.M. Vermaseren, S.A. Larin \& T. van Ritbergen, Phys. Lett.
{\bf B405} (1997), 327.
\bibitem{33} P.A. Baikov, K.G. Chetyrkin \& J.H. K\"{u}hn, JHEP {\bf 10} 
(2014), 076.
\bibitem{34} T. Luthe, A. Maier, P. Marquard \& Y. Schr\"{o}der, JHEP {\bf 01} 
(2017), 081.
\bibitem{35} P.A. Baikov, K.G. Chetyrkin \& J.H. K\"{u}hn, JHEP {\bf 04} 
(2017), 119.
\bibitem{36} T. Ueda, B. Ruijl \& J.A.M. Vermaseren, PoS (LL2016) (2016), 070.
\bibitem{37} T. Ueda, B. Ruijl \& J.A.M. Vermaseren, Comput. Phys. Commun.
{\bf 253} (2020), 107198.
\bibitem{38} F. Herzog, Nucl. Phys. {\bf B926} (2018), 370.
\bibitem{39} K.G. Chetyrkin, G. Falcioni, F. Herzog and J.A.M. Vermaseren, 
PoS (RADCOR2017) (2018), 004. 
\bibitem{40} B. Ruijl, T. Ueda, J.A.M. Vermaseren \& A. Vogt, JHEP {\bf 06} 
(2017), 040.
\bibitem{41} J.A.M. Vermaseren, math-ph/0010025.
\bibitem{42} M. Tentyukov \& J.A.M. Vermaseren, Comput. Phys. Commun. {\bf 181}
(2010), 1419.
\bibitem{43} S.A. Larin \& J.A.M. Vermaseren, Phys. Lett. {\bf B303} (1993),
334.
\bibitem{44} P. Nogueira, J. Comput. Phys. {\bf 105} (1993), 279.
\bibitem{45} T. van Ritbergen, A.N. Schellekens \& J.A.M. Vermaseren, Int. J.
Mod. Phys. {\bf A14} (1999), 41.
\bibitem{46} O.V. Tarasov, Phys. Part. Nucl. Lett. {\bf 17} (2020), 109.
\bibitem{47} B.A. Kniehl \& O.L. Veretin, Phys. Lett. {\bf B804} (2020),
13598.
\bibitem{48} Y.-J. Bi, H. Cai, Y. Chen, M. Gong, K.-F. Liu, Z. Liu \& Y.-B.
Yang, Phys. Rev. {\bf D97} (2018), 094501.
\end{thebibliography}
\end{document}